\documentclass[12pt]{article}
\usepackage{a4wide,graphicx,amsmath,amssymb,cite,longtable,multicol,array}
\usepackage{cancel}
\usepackage{comment}
\usepackage[usenames]{color}

\newcommand{\lsim}{\raisebox{-0.13cm}{~\shortstack{$<$ \\[-0.07cm] $\sim$}}~}
\newcommand{\gsim}{\raisebox{-0.13cm}{~\shortstack{$>$ \\[-0.07cm] $\sim$}}~}
\newcommand{\Raw}{\Rightarrow}
\newcommand{\nn}{\nonumber}
\newcommand{\bea}{\begin{eqnarray}}
\newcommand{\eea}{\end{eqnarray}}
\newcommand{\ba}{\begin{array}}
\newcommand{\ea}{\end{array}}
\newcommand{\dis}{\displaystyle}
\newcommand{\ii}{{\rm i}}

\newcommand{\e}{{\rm e}}

\def\slash#1{\setbox0=\hbox{$#1$}%
  \rlap{\ifdim\wd0>.7em\kern.22\wd0\else\kern.1\wd0\fi /}#1}

\makeatletter
\@addtoreset{equation}{section}

\makeatother

\newcommand{\muegamma}{\mu\to\e\gamma}
\newcommand{\mueee}{\mu\to\e\e\overline{\e}}
\newcommand{\mueN}{\mu\,{\rm N\to\e\, N}}

\begin{document}

\hfill\begin{tabular}{r}
CAFPE-135/10 \\
UG-FT-265/10\\
\end{tabular}

\bigskip
\bigskip
\bigskip
\bigskip

\begin{center}

\begin{Large}
\textbf{$\mu - \e$ conversion in the Littlest Higgs model \\[1ex]
        with T-parity}
\end{Large}

\bigskip
\bigskip

\begin{large}
F.~del~\'Aguila, 
J.~I.~Illana
and
M.~D.~Jenkins
\end{large}

\bigskip

\begin{it}
CAFPE and Departamento de F{\'\i}sica Te\'orica y del Cosmos, \\[1ex]
    Universidad de Granada, E--18071 Granada, Spain
\end{it}

\bigskip

{\tt faguila@ugr.es}, {\tt jillana@ugr.es}, {\tt mjenk@ugr.es}
\end{center}

\bigskip
\bigskip

\begin{abstract}
Little Higgs models provide a natural explanation of the little hierarchy between the electroweak scale and a few TeV scale, where new physics is expected. Under the same inspiring naturalness arguments, this work completes a previous study on lepton flavor-changing processes in the Littlest Higgs model with T-parity exploring the channel that will eventually turn out to be the most sensitive, $\mu-\e$ conversion in nuclei. All one-loop contributions are carefully taken into account, results for the most relevant nuclei are provided and a discussion of the influence of the quark mixing is included.
The results for the Ti nucleus are in good agreement with earlier work by Blanke et al., where a degenerate mirror quark sector was assumed. The conclusion is that, although this particular model reduces the tension with electroweak precision tests, if the restrictions on the parameter space derived from lepton flavor violation are taken seriously, the degree of fine tuning necessary to meet these constraints also disfavors this model.
\end{abstract}

\thispagestyle{empty}

\newpage
\tableofcontents

\newpage

\section{Introduction}

Lepton flavor is conserved within experimental limits in all known processes, 
except for neutrino oscillations. This is predicted by the Standard 
Model (SM) when it is minimally extended to include Dirac and/or Majorana masses \cite{Mohapatra:1998rq}. In such a case any neutral lepton flavor transition is proportional to the neutrino masses $m_\nu \sim 0.1$ eV, and then negligible 
if the relevant energy scale in the process $M$ is near or above the GeV. Hence, the observation of lepton flavor violation  (LFV) in processes like $\muegamma$, $\mueee$ or $\mueN$, or the analogous $\tau$ decays would be a clear signal of new physics beyond the SM. 

On the other hand, any new physics near the TeV scale involving leptons may 
induce large LFV transitions if it is not aligned with the SM charged lepton Yukawa couplings. In this case the relevant scales in the process are the electroweak scale $v \approx 246$ GeV and the new scale $f$ of few TeV, with the corresponding ratio $v / f \sim 0.1$ almost nine orders of magnitude larger than $m_\nu / M$.\footnote{
Note that amplitudes must be at least suppressed by one of these ratios, 
and then cross sections are at least quadratically suppressed.} Due to very stringent bounds on LFV processes (as small as $10^{-11}$ for $\muegamma$)
new physics at the TeV must incorporate a very efficient lepton flavor suppression mechanism to agree with current and foreseen limits on LFV transitions.  

Present limits on flavor violating processes set stringent constraints on the  
possible extensions of the SM, in particular on those solving 
the (little) hierarchy problem like supersymmetry, Little Higgs models and 
models with large extra dimensions. 
(See for a recent review \cite{Buras:2009if}, and references therein.) 
These bounds, which in the quark sector and for $\tau$ leptons have been improved by BaBar and Belle \cite{PDG}, are also expected to be improved for the first two lepton families in the near future. 
Thus MEG will improve the precision on $\muegamma$ and $\mueee$ by 
two orders of magnitude \cite{Ritt:2006cg,Mori:2007zza}, whereas PRISM/PRIME could reduce the bounds on 
$\mueN$ by several orders of magnitude \cite{Kuno:2005mm}. 
Super-B factories under study should also improve the precision on $\tau$ decays by an order of magnitude \cite{Bona:2007qt}.

Little Higgs models take care of the large top corrections to the Higgs mass by making the Higgs a pseudo-Goldstone boson of a larger global symmetry broken at a scale of a few TeV \cite{ArkaniHamed:2001ca,ArkaniHamed:2001nc, ArkaniHamed:2002qy,Schmaltz:2005ky,Han:2005ru,Perelstein:2005ka}.  This appealing solution to the little hierarchy problem encounters difficulties when a specific model is defined.  In particular, electroweak precision data \cite{Csaki:2002qg,Csaki:2003si,Han:2003wu,Kilian:2003xt,Hubisz:2005tx,Chen:2006dy} and the limits for flavor violating processes tend to banish the new physics scale to higher energies.  To deal with the electroweak data constraints, T-parity \cite{Cheng:2003ju,Cheng:2004yc,Low:2004xc} was introduced with the Littlest Higgs model with T-parity (LHT) being the most popular solution for a lighter new scale.

In this paper we re-evaluate $\mueN$ in the Littlest Higgs model with T-parity 
\cite{Blanke:2007db}. This complements our previous calculations of $\muegamma$ and $\mueee$ 
in \cite{delAguila:2008zu} and those in \cite{Choudhury:2006sq,Blanke:2007db}. 
The results for the Ti nucleus are in good agreement with previous work \cite{Blanke:2007db,Blanke:2009am} where a degenerate mirror
quark sector was assumed. Here we have also considered the effect of a more general quark sector where we include non-degenerate mirror quark masses and arbitrary mixings.
We have also found that $\mueN$ is one-loop finite \cite{Goto:2008fj,delAguila:2008zu,Blanke:2009am}. 
This is apparent for box diagrams but not for $Z$ penguins.
In general, this process provides a stronger bound on the new physics 
scale $f$, but limits from all three processes are required because there are accidental suppressions of any of them in certain regions of parameter space. Obviously, each of them cancel when the heavy lepton Yukawa couplings align with those of the charged leptons, that is the amplitudes vanish when the 
corresponding mixing angle $\theta$ goes to zero. 
The amplitudes also scale like $f^{-2}$ when the new physics scale goes to infinity with all other parameters finite. 
Present bounds on these processes are satisfied for $\theta \lsim 0.01$ or 
$f\gsim 10$ TeV. Particular cancellations result from amplitude contributions 
of different sign, for example between penguins and boxes, at given parameter values. As we shall discuss, in the case of $\mueN$ these cancellations occur even for degenerate quark masses (or just one quark family). 

On the other hand, box as well as $\gamma$ penguin contributions are at most 
logarithmically divergent when the internal fermion masses go to 
infinity, whereas the $Z$ penguins add a quadratically divergent 
contribution when these masses are parametrically increased. 
This is similar to what happens in the SM with the top mass $m_t$ \cite{Sirlin:1980nh,Hollik:1988ii} (and to $\mueee$ in these models \cite{delAguila:2008zu}). 
In our case the heavy fermion masses are proportional to $\kappa f$, 
where $\kappa$ is the corresponding Yukawa coupling. 
For a fixed $f$ value, the $Z$ penguins scale like $\kappa^2$, the same as, for instance, the top quark contribution to $Z b\bar b$ that scales like $\lambda^2$, with $m_t\sim\lambda v$ \cite{Bernabeu:1987me}. 
In both cases the gauge symmetry, 
$[{\rm SU}(2)\times{\rm U}(1)]_1\times[{\rm SU}(2)\times{\rm U}(1)]_2$ 
in the LHT and ${\rm SU}(2)_L\times{\rm U}(1)_Y$ in the SM, 
is spontaneously broken and the heavy fermion mass splits the fermion multiplet.     
Of course, these Yukawa couplings can be large and the 
corresponding contributions dominant, but they should eventually remain perturbative. 

In a companion paper we present the evaluation of the three processes $\muegamma$, $\mueee$ and $\mueN$ in the Simplest Little Higgs model \cite{paper3}, finding similar results \cite{Illana:2009yp}. These calculations can also be applied to the four-dimensional formulation of extended models with large extra dimensions \cite{Agashe:2006iy,delAguila:2010vg}. 
In this case, however, care must be taken to include the appropriate 
number of Kaluza-Klein modes to match the five-dimensional result \cite{Csaki:2010aj}. 
 
In Section 2 the LHT model is briefly described with the purpose of introducing notation and conventions, providing the Feynman rules for quarks, needed to perform the calculation of $\mueN$ 
which were not included in \cite{delAguila:2008zu}. (They agree with those in \cite{Blanke:2006eb} up to sign conventions.) 
In Section 3 we detail the different contributions, finding complete agreement with 
\cite{Blanke:2007db} in the total sum. 
We show in Section 4 that, in general, the conversion rates for both Ti and Au give a more stringent limit than $\muegamma$ and $\mueee$, with Au being the most stringent. Results for Pb are not shown but it was found that it provides a lower conversion rate than Ti. We also study the effect of the heavy quark sector on the conversion rate. A degenerate heavy quark sector was previously assumed 
in the literature arguing that the quark mixing is a higher order effect. 
We analyze the degenerate case, and then show the impact of a non-degenerate heavy quark sector including quark mixings. Even assuming degeneracy, the quark mass parameter can entirely cancel the conversion rate if chosen appropriately (in a region where box diagrams cancel penguin diagrams). Although the effect of the mixings is found to be less important than that of the quark masses, they can have sizable effects on the conversion rate since they can push the values of the form factors into regions with cancellations. 
Finally, Section 5 is devoted to our conclusions. 

\section{The LHT model}

The Littlest Higgs \cite{ArkaniHamed:2002qy} is a non-linear $\sigma$ model based on the spontaneous breaking of a global SU(5) symmetry into SO(5) by the vacuum expectation value of a five-dimensional tensor field. The 14 Goldstone fields are parameterized as
\bea
\Sigma(x)=\xi^2\Sigma_0,\quad
\xi={\rm e}^{\ii\Pi(x)/f},\quad 
\Sigma_0=\left(\ba{ccc} {\bf 0}_{2\times2} & 0 & {\bf 1}_{2\times2} \\
                         0 & 1 &0 \\ 
                         {\bf 1}_{2\times2} & 0 & {\bf 0}_{2\times 2}\ea\right).
\label{GT}
\eea
Only an $[{\rm SU}(2)\times{\rm U}(1)]_1\times[{\rm SU}(2)\times{\rm U}(1)]_2$ subgroup of SU(5) is gauged, with generators
\bea
Q_1^a=\frac{1}{2}\left(\ba{ccc}\sigma^a & 0 & 0 \\ 0 & 0 & 0 \\ 0 & 0 &  {\bf 0}_{2\times2} \ea\right),\quad&&
Y_1=\frac{1}{10}{\rm diag}(3,3,-2,-2,-2),
\\
Q_2^a=\frac{1}{2}\left(\ba{ccc}{\bf 0}_{2\times2} & 0 & 0 \\ 0 & 0 & 0 \\ 0 & 0 & -\sigma^{a*} \ea\right),\quad&&
Y_2=\frac{1}{10}{\rm diag}(2,2,2,-3,-3),
\eea
and $\sigma^a$ the three Pauli matrices. This gauge group is broken by $\Sigma_0$ down to the SM group, whose generators are the combinations $\{Q_1^a+Q_2^a,\ Y_1+Y_2\}$. A discrete T-symmetry is introduced \cite{Cheng:2003ju,Cheng:2004yc,Low:2004xc}
that exchanges the gauge fields $G_{1,2}$ of $[{\rm SU}(2)\times{\rm U}(1)]_{1,2}$ under the assumption that $g\equiv g_1=g_2$, $g'\equiv g'_1=g'_2$. The T-even combinations remain massless while those T-odd acquire a mass proportional to $f$. In order that the SM Higgs doublet contained in $\Pi$ is T-even and the remaining Goldstone fields T-odd, the T transformation on the scalar fields is defined as follows,
\bea
\Pi\stackrel{\rm T}{\longrightarrow}-\Omega\Pi\Omega,\quad
\Omega={\rm diag}(-1,-1,1,-1,-1)
\eea
and then
\bea
\Sigma\stackrel{{\rm T}}{\longrightarrow}
\widetilde\Sigma=\Omega\Sigma_0\Sigma^\dagger\Sigma_0\Omega,\quad
\xi\stackrel{{\rm T}}{\longrightarrow}\Omega\xi^\dagger\Omega.
\eea

The gauge- and T-invariant Lagrangians for the vector and Goldstone bosons can then be written in terms of covariant derivatives. Our sign conventions are those in \cite{delAguila:2008zu}. 
After the electroweak symmetry breaking the photon remains massless, the weak gauge bosons pick up a mass of order $v$, the T-odd gauge bosons get rotated to their physical states $A_H$, $Z_H$ and $W_H^\pm$, and the scalar fields are also shifted and rotated accordingly. The latter include the would-be Goldstone fields $\eta$, $\omega^0$, $\omega^\pm$ eaten by the four heavy gauge bosons above. 

The introduction of fermions in the model is less straightforward. Since the implementation of the lepton sector was described in detail in \cite{delAguila:2008zu}, which was sufficient for the processes $\muegamma$ and $\mueee$, we present below the relevant Yukawa and gauge interaction Lagrangians for the quark sector, also needed to study $\mueN$.

\subsection{Quark Lagrangians}

As for the lepton sector, one introduces two left-handed fermion doublets in incomplete SU(5) multiplets, one transforming only under SU(2)$_1$ and the other under SU(2)$_2$, for each SM left-handed quark doublet \cite{Low:2004xc}: 
\begin{equation}
 \Psi_1^{[{\bf \bar 5}]} = \left( \begin{array}{c} -\ii \sigma^2 q_{1L} \\ 0 \\ 0 \end{array} \right), \quad
 \Psi_2^{[{\bf 5}]}  = \left( \begin{array}{c} 0 \\ 0 \\-\ii \sigma^2 q_{2L}\end{array} \right),
\end{equation}
where $q_{iL} = \left( \begin{array}{c} u_{iL} \\ d_{iL} \end{array} \right)$.
An SU(5) transformation $V$ and a T transformation acts as follows:
\begin{equation}
 \Psi_1 \longrightarrow V^*\Psi_1, \quad 
 \Psi_2 \longrightarrow V\Psi_2, 
 \quad \Psi_1 \stackrel{\rm T}{\longleftrightarrow} \Omega \Sigma_0 \Psi_2.
\label{psitrans}
\end{equation}
The T-even combination $\Psi_1 + \Omega \Sigma_0 \Psi_2$ remains massless
while the T-odd $\Psi_1 - \Omega \Sigma_0 \Psi_2$ obtains a mass when coupled (see below) to a multiplet $\Psi_R$ defined as
\begin{equation}
 \Psi_R = \left( \begin{array}{c} \tilde{\psi}_R \\ \chi_R \\ -\ii \sigma^2 q_{HR} \end{array} \right), \quad q_{HR} = \left(\begin{array}{c} u_{HR} \\ d_{HR} \end{array}\right), \quad 
\Psi_R \longrightarrow U\Psi_R, \quad 
\Psi_R \stackrel{\rm T}{\longrightarrow}\Omega \Psi_R,
\end{equation}
so that $\xi\Psi_R$ transforms as a {\bf 5} representation.
The operator $U$ is a non-linear transformation depending on $V$ and the Goldstone fields, that takes values in the Lie algebra of the unbroken SO(5), defined by
\bea
\Sigma\equiv\xi^2\Sigma_0\longrightarrow V\Sigma V^T \quad 
\Raw \quad \xi\longrightarrow V\xi U^\dagger \equiv U\xi\Sigma_0 V^T\Sigma_0,
\eea
such that the following Yukawa Lagrangian is SU(5) and T invariant,
\bea
\mathcal{L}_{Y_H} = -\kappa f \left(\bar{\Psi}_2\xi\Psi_R + \bar{\Psi}_1\Sigma_0\xi^\dag\Psi_R\right) + \textrm{h.c.}\ .
\label{YH}
\eea
Defining the T-eigenstates
\begin{equation}
 q_L = \frac{q_{1L} - q_{2L}}{\sqrt{2}}, \quad q_{HL} = \frac{q_{1L} + q_{2L}}{\sqrt{2}}, \quad q=u,d,
\end{equation}
we see that the left-handed T-odd fields $u_{HL}$ and $d_{HL}$ couple to $u_{HR}$ and $d_{HR}$, respectively, to get masses of order $f$. The T-even combinations $u_L$ and $d_L$ are the SM left-handed quarks.
The fields $\chi_R$ and $\tilde\psi_R$ can be assumed to acquire large Dirac masses pairing with additional fermions \cite{Low:2004xc,Hubisz:2004ft} so that they decouple from the theory. 

The left-handed quark fields interact with the gauge bosons through the following covariant derivatives:
\bea
\mathcal{L}_{FL}&=&\ii \bar{\Psi}_1 \cancel{D}_{1}\Psi_1 + \ii\bar{\Psi}_2\cancel{D}_{2}\Psi_2, \label{FGHL}
\eea
where
\bea
D_{1\mu}&=&\partial_{\mu}+\sqrt{2}\ii g(W_{1\mu}^aQ_{1}^{aT}+W_{2\mu}^aQ_2^{aT})+
\sqrt{2}\ii g'(y_1^{\Psi_1}B_{1\mu}+y_2^{\Psi_1}B_{2\mu}),\\
D_{2\mu}&=&\partial_{\mu}-\sqrt{2}\ii g(W_{1\mu}^aQ_{1}^{a}+W_{2\mu}^aQ_2^{a})+
\sqrt{2}\ii g'(y_1^{\Psi_2}B_{1\mu}+y_2^{\Psi_2}B_{2\mu}),
\end{eqnarray}
with $W_{j\mu}^a$, $B_{j\mu}$ the gauge fields of $[{\rm SU}(2)\times{\rm U}(1)]_j$ and
\bea
y_1^{\Psi_1}=y_2^{\Psi_2}=\frac{1}{30},\quad
y_2^{\Psi_1}=y_1^{\Psi_2}=\frac{2}{15} .
\eea
It is necessary to enlarge SU(5) with two extra U(1) factors to introduce these hypercharges \cite{Goto:2008fj}, that add up to $1/6$ (the hypercharge of the SM left-handed quark doublet).

The gauge interactions of the heavy and light right-handed fields are not necessary for our calculation since they turn out to be of higher order in $v/f$, but we include them for reference \cite{Hubisz:2004ft}:
\begin{eqnarray}
 \mathcal{L}_{FR} & = & \ii \overline{\Psi}_R\gamma^\mu\left[\partial_\mu + \frac{1}{2} \xi^\dag(D_{2\mu} \xi) + \frac{1}{2} \xi(\Sigma_0 D_{2\mu}^* \Sigma_0 \xi^\dag) \right] \Psi_R, \\
 \mathcal{L}'_{FR} & = & \ii \overline{u}_R \gamma^\mu(\partial_\mu + \ii g' y_u B_\mu) u_R + \ii \overline{d}_R \gamma^\mu(\partial_\mu + \ii g' y_d B_\mu) d_R,
\end{eqnarray}
where $y_u=2/3$, $y_d=-1/3$ are hypercharges that also require the two additional U(1) factors \cite{Goto:2008fj}.

In the case of three generations, the Yukawa Lagrangian (\ref{YH}) that provides the masses for the heavy quark states is given by:
\begin{equation}
\mathcal{L}_{Y_H} = -\kappa_{mn} f \left(\bar{\Psi}^m_2\xi\Psi_R^n + \bar{\Psi}^m_1\Sigma_0\xi^\dag\Psi_R^n\right) + \textrm{h.c.},
\end{equation}
with family indices $m,n= 1,2,3$.  This is the primary source of heavy flavor mixing in the quark sector, which is completely analogous to the lepton sector.  After diagonalization, ${\rm diag}(\kappa_m)=V_L^{H\dag}\kappa V_R^H$ where $V_L^H$ ($V_R^H$) acts on the left (right) handed fields, one finds that the masses for the heavy quarks are approximately given by:
\begin{equation}
 m_{d_H^m} = \sqrt{2}\kappa_m f, \quad m_{u_H^m} = \sqrt{2}\kappa_m f \left(1 - \frac{v^2}{8f^2} \right).
\label{mH}
\end{equation}
Then the T-odd gauge boson interactions from (\ref{FGHL}) work out as:
\begin{equation}
 \overline{q}_{HL} V_L^{H\dag} \cancel{G}_H \left( \begin{array}{c} V_L^u u_L \\ V_L^d d_L \end{array} \right) + \textrm{h.c.},
\end{equation}
where $G_H$ is a T-odd gauge boson ($A_H,Z_H,W_H$) and the matrices $V_L^{u,d}$ are rotations necessary to make the light sector mass diagonal. The matrices appearing in the coupling can therefore be defined as \cite{Hubisz:2005bd}:
\begin{equation}
 V_{Hu} \equiv V_L^{H\dag}V_L^u, \quad V_{Hd} \equiv V_L^{H\dag} V_L^d.
\end{equation}
The two light rotations are then related by $V_L^{u\dag}V_L^d = V_{\rm CKM}$ and can be obtained from the Yukawa Lagrangians for the light sector which we now describe.

For the SM quark Yukawa couplings, the first two generations are treated separately from the third one.  This is because of the special structure necessary for the collective symmetry breaking mechanism.  The third generation requires enlarging $\Psi_1$ and $\Psi_2$ into multiplets of the SU(3)$_1$ and SU(3)$_2$ subgroups of SU(5) \cite{Hubisz:2005bd} adding additional T-even and T-odd partners (even and odd combinations of $t'_1$ and $t'_2$), usually denoted by $T_+$ and $T_-$ respectively,
\begin{eqnarray}
Q_1^3=\left(\ii d_{1L}^3 , -\ii u_{1L}^3 , t'_{1L} , 0 , 0 \right)^T, \qquad
Q_2^3=\left(0 , 0 , t'_{2L} , \ii d_{2L}^3 , -\ii u_{2L}^3 \right)^T.
\end{eqnarray}
They have the same transformation properties as $\Psi_{1,2}$ in (\ref{psitrans}).
The quarks masses of this generation originate from the Lagrangian
\begin{equation}
\mathcal{L}_{Y_t} = -\frac{\lambda_1}{2\sqrt{2}}f\epsilon_{ijk}\epsilon_{xy} \left[ \bar{Q}_{1i}^3 \Sigma_{jx}\Sigma_{ky} - \left(\bar{Q}_2^3\Sigma_0\right)_i \tilde{\Sigma}_{jx} \tilde{\Sigma}_{ky} \right] u_R^3 + \lambda_2 f (\bar{t}'_{1L}t'_{1R} + \bar{t}'_{2L} t'_{2R} ) + \textrm{h.c.},
\end{equation}
where $i,j,k=1,2,3$ and $x,y=4,5$.

The first two generations are implemented differently and do not require the enlarged SU(3) multiplets:
\begin{equation}
\mathcal{L}_{Y_u}=-\frac{\lambda_u^{m}}{2\sqrt{2}}f\epsilon_{ijk}\epsilon_{xy}\left[\bar{Q}_{1i}^m\Sigma_{jx}\Sigma_{ky}-
\left(\bar{Q}^m_2\Sigma_0\right)_i \tilde{\Sigma}_{jx}\tilde{\Sigma}_{ky}\right]u_R^m
+ \textrm{h.c.} \quad (m=1,2),
\end{equation}
where $i,j,k=1,2,3$; $x,y=4,5$ and
\begin{eqnarray}
Q_1^m=\left(\ii d_{1L}^m , -\ii u_{1L}^m , 0 , 0 , 0 \right)^T, \qquad
Q_2^m=\left(0 , 0 , 0 , \ii d_{2L}^m , -\ii u_{2L}^m \right)^T \quad (m=1,2).
\end{eqnarray}

Unlike the up quark case, all three generations of down-type light quarks acquire mass through the same type of Lagrangian.  For simplicity, we have introduced the quark mixing (CKM matrix) in this sector only, choosing the up-type quarks to be directly in the mass basis.  This is because of the fact that the three generations of up quarks are implemented differently, which explicitly breaks flavor symmetries.  Although it is possible to restore flavor symmetries (see \cite{Hubisz:2005bd} for further details), this is unimportant in our case and it is sufficient to assume that all the mixing comes from the rotation in the down sector. Then
\begin{equation}
\mathcal{L}_{Y_d}=\frac{\ii\lambda^{mn}_d}{2\sqrt{2}}f\epsilon_{ij}\epsilon_{xyz}\left[\bar{\Psi'}^m_{2x}\Sigma_{iy}\Sigma_{jz}X-
\left(\bar{\Psi'}_1^m\Sigma_0\right)_x \tilde{\Sigma}_{iy}\tilde{\Sigma}_{jz}\tilde{X}\right]d_R^n 
+ \textrm{h.c.} \quad (m,n=1,2,3),
\end{equation}
where $i,j=1,2$; $x,y,z=3,4,5$, $X=(\Sigma_{33})^{-1/4}$ and
\begin{eqnarray}
{\Psi'}_1^m=\left(u_{1L}^m , d_{1L}^m , 0 , 0 , 0 \right)^T, \qquad
{\Psi'}_2^m=\left(0 , 0 , 0 , u_{2L}^m , d_{2L}^m \right)^T.
\end{eqnarray}
In this context, the previously defined $V_L^u$ is set to the identity matrix since there is no family mixing in the up sector and $V_L^d$ would correspond to $V_{\rm CKM}$ and would be one of the two matrices necessary to diagonalize $\lambda^{mn}_d$.  The other matrix, $V_R^d$ would be the rotation for the right handed sector.

The relevant degrees of freedom in $V_{Hu}$ (or $V_{Hd}$) are three angles, $\theta^u_{12}$, $\theta^u_{23}$ and $\theta^u_{13}$, and 
{\em three} complex phases, as emphasized in \cite{Blanke:2006xr} where a standard parameterization is proposed. 

\subsection{Feynman rules for the quark sector}

From these Lagrangians one can obtain the Feynman rules for the quark vertices entering in our calculation.  They are in agreement with \cite{Blanke:2006eb} and are summarized in tables~\ref{tab1} and \ref{tab2}, in our sign conventions.  Standard Model Feynman rules are used for vertices that involve only ordinary particles and they will not be listed.  Although the model predicts some corrections to some of these standard vertices, their contributions are subleading in our process and we can safely ignore them.

\begin{table}
\begin{center}
\begin{tabular}{|c|c|c|}
\hline
VFF & $g_L$ & $g_R$ \\
\hline
$A_H\bar u_H^i u^j$ & $\left(\dis\frac{1}{10c_W}+\frac{x_H}{2s_W}\frac{v^2}{f^2}\right)V_{Hu}^{ij}$ & 0 \\ \hline
$Z_H\bar u_H^i u^j$ & $\left(\dis\frac{1}{2s_W}-\frac{x_H}{10c_W}\frac{v^2}{f^2}\right)V_{Hu}^{ij}$ & 0 \\ \hline
$W_H^+\bar u_H^i d^j$ & $\dis\frac{1}{\sqrt{2}s_W}V_{Hd}^{ij}$ & 0\\ \hline
$A_H\bar d_H^i d^j$ & $\left(\dis\frac{1}{10c_W}-\frac{x_H}{2s_W}\frac{v^2}{f^2}\right)V_{Hd}^{ij}$ & 0 \\ \hline
$Z_H\bar d_H^i d^j$ & $-\left(\dis\frac{1}{2s_W}+\frac{x_H}{10c_W}\frac{v^2}{f^2}\right)V_{Hd}^{ij}$ & 0 \\ \hline
$W_H^-\bar d_H^i u^j$ & $\dis\frac{1}{\sqrt{2}s_W}V_{Hu}^{ij}$ & 0 \\ \hline
\end{tabular}
\end{center}
\caption{VFF vertices $\ii e\gamma^\mu(g_LP_L+g_RP_R)$ for quarks in the LHT model; $x_H=\dis\frac{5gg'}{4(5g^2-g'^2)}.$\label{tab1}}
\end{table}

\begin{table}
\begin{center}
\begin{tabular}{|c|c|c|}
\hline
SFF & $c_L$ & $c_R$ \\
\hline
$\eta\bar u_H^i u^j$ & $\dis\frac{\ii}{10c_W}\frac{m_{u_H^i}}{M_{A_H}}\left[1+\frac{v^2}{f^2}
\left(\frac{5}{8}+x_H\frac{s_W}{c_W}\right)\right]
V_{Hu}^{ij}$ & $-\dis\frac{\ii}{10c_W}\frac{m_{u^i}}{M_{A_H}}V_{Hu}^{ij}$ \\ \hline
$\omega^0\bar u_H^i u^j$ & $-\dis\frac{\ii}{2s_W}\frac{m_{u_H^i}}{M_{Z_H}}\left[1+\frac{v^2}{f^2}
\left(\frac{1}{8}-x_H\frac{c_W}{s_W}\right)\right]
V_{Hu}^{ij}$ & $\dis\frac{\ii}{2s_W}\frac{m_{u^i}}{M_{Z_H}}V_{Hu}^{ij}$  \\ \hline
$\omega^+\bar u_H^i d^j$ & $-\dis\frac{\ii}{\sqrt{2}s_W}\frac{m_{u_H^i}}{M_{W_H}}
V_{Hd}^{ij}$ & $\dis\frac{\ii}{\sqrt{2}s_W}\frac{m_{d^i}}{M_{W_H}}V_{Hd}^{ij}$ \\ \hline
$\eta\bar d_H^i d^j$ & $\dis\frac{\ii}{10c_W}\frac{m_{d_H^i}}{M_{A_H}}\left[1-\frac{v^2}{f^2}
\left(\frac{5}{4}+x_H\frac{s_W}{c_W}\right)\right]
V_{Hd}^{ij}$ & $-\dis\frac{\ii}{10c_W}\frac{m_{d^i}}{M_{A_H}}V_{Hd}^{ij}$\\ \hline
$\omega^0\bar d_H^i d^j$ & $\dis\frac{\ii}{2s_W}\frac{m_{d_H^i}}{M_{Z_H}}\left[1+\frac{v^2}{f^2}
\left(-\frac{1}{4}+x_H\frac{c_W}{s_W}\right)\right]
V_{Hd}^{ij}$  & $-\dis\frac{\ii}{2s_W}\frac{m_{d^i}}{M_{Z_H}}V_{Hd}^{ij}$ \\ \hline
$\omega^-\bar d_H^i u^j$ & $-\dis\frac{\ii}{\sqrt{2}s_W}\frac{m_{d_H^i}}{M_{W_H}}\left(1-\frac{v^2}{8f^2}\right)V_{Hu}^{ij}$ & $\dis\frac{\ii}{\sqrt{2}s_W}\frac{m_{u^i}}{M_{W_H}}V_{Hu}^{ij}$ 
\\ \hline
\end{tabular}
\end{center}
\caption{SFF vertices $\ii e(c_LP_L+c_RP_R)$ for quarks in the LHT model; $x_H=\dis\frac{5gg'}{4(5g^2-g'^2)}.$ \label{tab2}}
\end{table}

\section{Contributions to the $\mueN$ process}

\subsection{Topologies and form factors}

\begin{figure}
 \begin{center}
\begin{tabular}{cc}
\includegraphics[width=4.5cm]{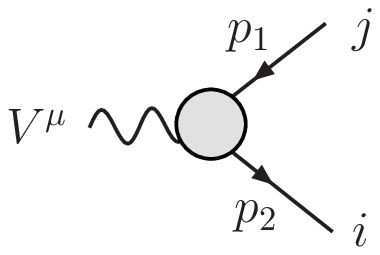} &
\includegraphics[width=4.5cm]{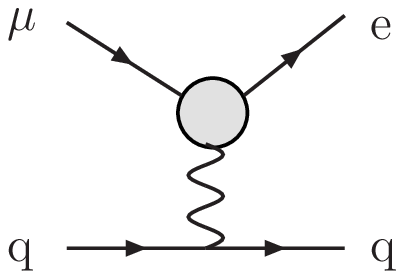} 
\includegraphics[width=4.5cm]{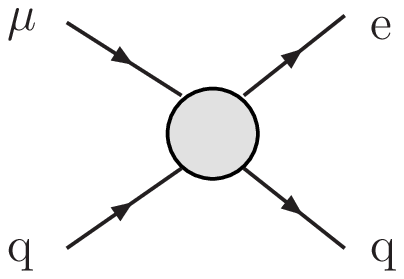} \\
(a) & (b)
\end{tabular}
\end{center}
\caption{Vertex subdiagrams (a) and topologies (b) involved in $\mueN$.}
\label{topologies}
\end{figure}

The $\mueN$ process receives contributions from both penguin and box diagrams like those shown in figure~\ref{topologies}b.  Here the quark $q$ is either an up or down quark and is of the same type before and after the interaction.  The gauge bosons exchanged in the penguin diagrams are a Z boson or a photon.  The diagrams look very similar to those involved in $\mueee$ except for the fact that now up and down-type quarks instead of electrons are involved on the lower legs.

The contributions of the lepton flavor-changing vertex subdiagrams fit into the following Lorentz structure \cite{Hisano:1995cp,Illana:2000ic,Illana:2002tg}:\footnote{
A minus sign has been extracted from the dipole form factors $F_M$ and $F_E$ in order to meet the definition of the momentum transfer $Q$ in \cite{Hisano:1995cp}, that had opposite sign in \cite{Illana:2000ic,Illana:2002tg}. This implies a sign flip in (\ref{ffcom}) accordingly.}
\begin{eqnarray}
{\rm i}\Gamma^\mu (p_1,p_2) = {\rm i} e\left[ \gamma^\mu(F_L^V P_L + F_R^V P_R) - ({\rm i} F_M^V + F_E^V \gamma^5)\sigma^{\mu\nu}Q_\nu + ({\rm i }F_S^V + F_P^V \gamma^5)Q^\mu \right], \label{trianglesff}
\end{eqnarray}
where $V$ denotes the external vector boson, either Z or $\gamma$ in our case, and $Q = p_1 - p_2$ as in figure~\ref{topologies}a. The penguin diagrams for $\mu\to{\rm eq\bar q}$ can be read from \cite{delAguila:2008zu} introducing the corresponding electric charges ($Q_q$) and weak couplings ($Z_{L,R}^q$) for quarks:
\bea
\mathcal{M}_{\gamma \textrm{peng}}&=&-\frac{e^2}{Q^2}\bar{u}(p_1)
\left[Q^2\gamma^\mu(A_1^LP_L+A_1^RP_R)+\ii m \sigma^{\mu\nu}Q_\nu(A_2^LP_L+A_2^RP_R)\right]u(p)\nonumber \\
&&\times \bar{u}_q(p_2)Q_q\gamma_\mu v_q(p_3),
\label{e1}\\
\mathcal{M}_{\rm Zpeng} &=& \frac{e^2}{M^2_Z}\bar{u}(p_1)
\left[\gamma^\mu(F_LP_L+F_RP_R)\right]u(p)\bar{u}_q(p_2)[\gamma_\mu (Z_L^qP_L+Z_R^qP_R)]v_q(p_3),
\label{e2}
\eea
where $m$ is the muon mass, the electron mass has been neglected, and
\bea
&&A_1^L=F_L^\gamma/Q^2, \
  A_1^R=F_R^\gamma/Q^2, \
  A_2^L=-(F_M^\gamma+\ii F_E^\gamma)/m, \
  A_2^R=-(F_M^\gamma-\ii F_E^\gamma)/m, \nn\\
&&F_L=-F_L^Z, \quad
  F_R=-F_R^Z.
\label{ffcom}
\eea
The form factors $F_S$ and $F_P$ do not contribute in any case since they are found to be proportional to the ratio of the external lepton and quark masses to the mass of the W boson and are therefore negligible.  The same occurs in the case of $F_E$ and $F_M$ in the Z boson penguin.

Similarly, in the limit of vanishing external momenta the contribution to the amplitude from any box diagram can be written as \cite{Hisano:1995cp}:
\begin{eqnarray}
{\cal M}_{\rm box}^q&=&
\quad e^2 B_{1q}^L\left[\bar u(p_1)\gamma^\mu P_L u(p)\right]
            \left[\bar u_q(p_2)\gamma_\mu P_L v_q(p_3)\right] \nn\\
&&+e^2 B_{1q}^R\left[\bar u(p_1)\gamma^\mu P_R u(p)\right]
            \left[\bar u_q(p_2)\gamma_\mu P_R v_q(p_3)\right] \nn\\
&&+e^2 B_{2q}^L\left[\bar u(p_1)\gamma^\mu P_L u(p)\right]
            \left[\bar u_q(p_2)\gamma_\mu P_R v_q(p_3)\right] \nn\\
&&+e^2 B_{2q}^R\left[\bar u(p_1)\gamma^\mu P_R u(p)\right]
            \left[\bar u_q(p_2)\gamma_\mu P_L v_q(p_3)\right] \nn\\
&&+e^2 B_{3q}^L\left[\bar u(p_1) P_L u(p)\right]
            \left[\bar u_q(p_2) P_L v_q(p_3)\right] \nn\\
&&+e^2 B_{3q}^R\left[\bar u(p_1) P_R u(p)\right]
            \left[\bar u_q(p_2) P_R v_q(p_3)\right] \nn\\
&&+e^2 B_{4q}^L\left[\bar u(p_1) \sigma^{\mu\nu}P_L u(p)\right]
            \left[\bar u_q(p_2)\sigma_{\mu\nu} P_L v_q(p_3)\right] \nn\\
&&+e^2 B_{4q}^R\left[\bar u(p_1) \sigma^{\mu\nu}P_R u(p)\right]
            \left[\bar u_q(p_2)\sigma_{\mu\nu} P_R v_q(p_3)\right].\label{bfact}
\label{boxesff}
\end{eqnarray}
However, we may restrict ourselves to the first term, proportional to $B_{1q}^L$, since it is the only contributing due to the fact that the LHT couplings are primarily left-handed. 

Finally, in the process $\mueN$ only the quark vector current contributes \cite{Kuno:1999jp}. The conversion width is then given by \cite{Hisano:1995cp}:
\begin{equation}
\Gamma(\mu\to \textrm{e})=\alpha^5\frac{Z_{\rm eff}^4}{Z}F(q)^2m_{\mu}^5 
\left| 2Z(A_1^L-A_2^R)-(2Z+N)\bar{B}_{1u}^L-(Z+2N)\bar{B}_{1d}^L\right|^2,
\end{equation}
where
\begin{equation}
\bar{B}_{1q}^L=B_{1q}^L+\frac{F_L}{M_Z^2}(Z_L^q+Z_R^q)\equiv B_{1q}^L+F_{LL}^q+F_{LR}^q.
\end{equation}
The conversion rate is obtained by dividing by the capture rate:
\begin{equation}
\mathcal{R}=\frac{\Gamma(\mu\to {\rm e})}{\Gamma_{\rm capt}}.
\end{equation}
The nuclei we will consider are ${}^{48}_{22}\textrm{Ti}$ and ${}^{197}_{\ 79}\textrm{Au}$, whose relevant parameters are listed in table~\ref{tab3}. 
We now calculate the form factors appearing in the amplitudes in terms of standard loop functions.

\begin{table}
\begin{center}
\begin{tabular}{|c|rrccc|}
\hline
Nucleus & $Z$ & $N$ & $Z_{\rm eff}$ & $F(q)$ & $\Gamma_{\rm capt}$ [GeV] \\
\hline
${}^{48}_{22}\textrm{Ti}$ & 22 & 26 & 17.6 & 0.54 &  $1.7\times 10^{-18}$ \\
${}^{197}_{\ 79}\textrm{Au}$ & 79 & 118 & 33.5 & 0.16 & $8.6\times 10^{-18}$ \\
\hline
\end{tabular}
\end{center}
\caption{Relevant input parameters for the nuclei under study. From \cite{Kitano:2002mt}. \label{tab3}}
\end{table}

\subsection{Penguin diagrams}

For the penguin diagrams, the contributions to the form factors are identical to the $\mueee$ case since they only depend on the lepton triangle diagrams.  The only difference here comes from the couplings to quarks and these are taken into account in the expressions (\ref{e1}) and (\ref{e2}) for the matrix elements.  We therefore just recall our previous results:\footnote{
The sign of the dipole form factor $A_2^R$ has been reversed as compared to \cite{delAguila:2008zu}, where the sign of $Q$ was mistaken. This has no consequences in $\muegamma$ and a non-significant impact on $\mueee$.}
\begin{eqnarray}
A_1^L & = & \frac{\alpha_W}{4\pi}\frac{1}{M_W^2}\frac{v^2}{4f^2}
	\sum_i V_{H\ell}^{ie*}V_{H\ell}^{i\mu} 
	\Big[ G_W^{(1)}(y_i)+G_Z^{(1)}(y_i)+\frac{1}{5}G_Z^{(1)}(ay_i)\Big],\\
A_2^R & = & -\frac{\alpha_W}{8\pi}\frac{1}{M_W^2}\frac{v^2}{4f^2}
	\sum_i V_{H\ell}^{ie*}V_{H\ell}^{i\mu} 
	\Big[ F_W(y_i)+F_Z(y_i)+\frac{1}{5}F_Z(ay_i)\Big],\\
F_L  & = & -\frac{\alpha_W}{8\pi}\frac{1}{s_Wc_W}\frac{v^2}{8f^2}
	\sum_i V_{H\ell}^{ie*}V_{H\ell}^{i\mu} \, y_i H_W(y_i),
\label{Fpenguin}
\end{eqnarray}
where 
\bea
y_i = \frac{m_{Hi}^2}{M_{W_H}^2},\quad
a= \frac{M_{W_H}^2}{M_{A_H}^2} = \frac{5c_W^2}{s_W^2},
\eea
and the loop functions are given in equations (3.23), (3.29), (3.35), (3.39) and (3.42) of \cite{delAguila:2008zu}.

\subsection{Box diagrams}

\begin{figure}[p]
\centering
\begin{tabular}{cccc}
\includegraphics[scale=0.62]{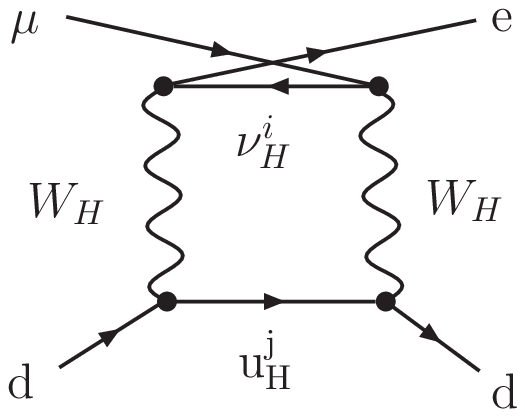}   & \hspace{-8mm}
\includegraphics[scale=0.62]{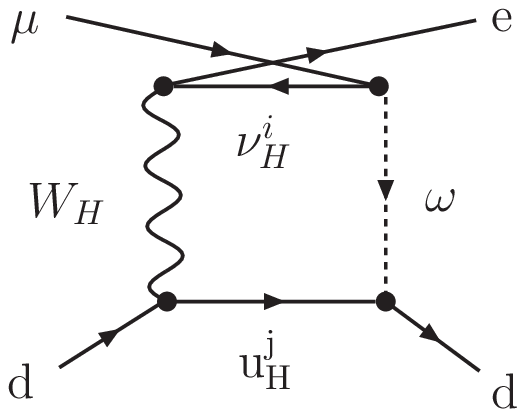}   & \hspace{-8mm}
\includegraphics[scale=0.62]{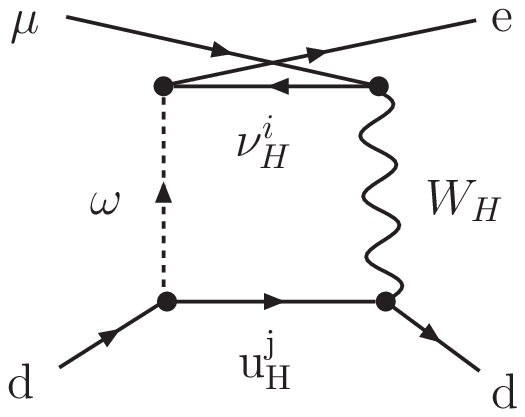}   & \hspace{-8mm}
\includegraphics[scale=0.62]{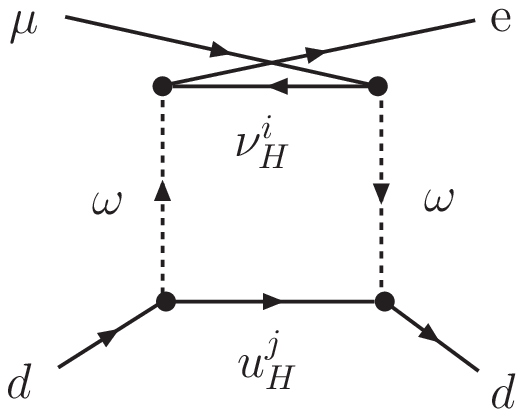}     \hspace{-8mm}
\\
\includegraphics[scale=0.62]{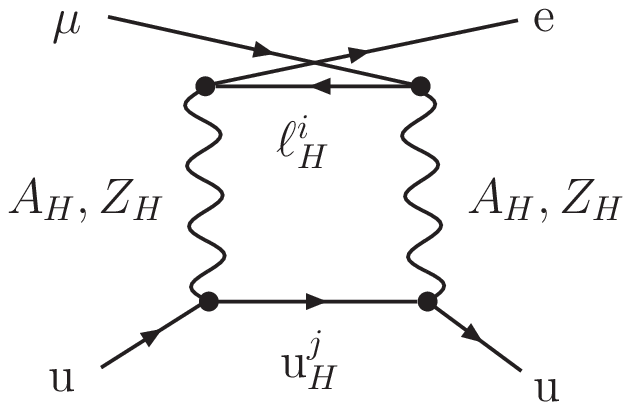} & \hspace{-8mm}
\includegraphics[scale=0.62]{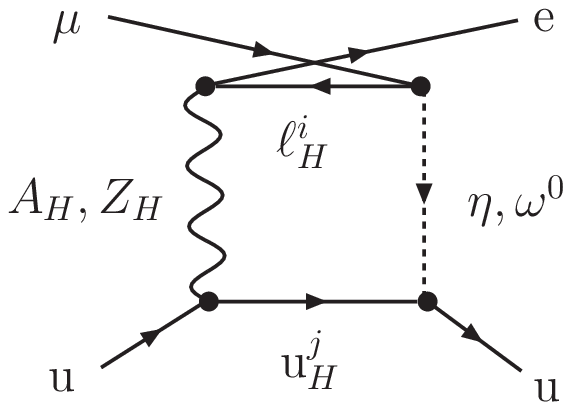} & \hspace{-8mm}
\includegraphics[scale=0.62]{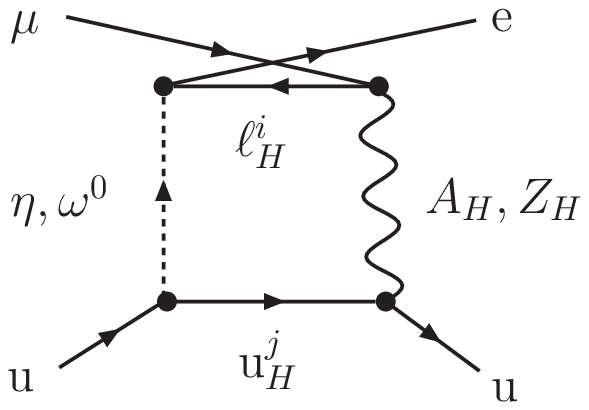} & \hspace{-8mm}
\includegraphics[scale=0.62]{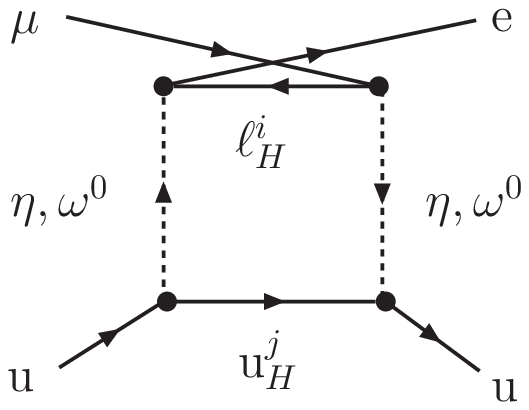}   \hspace{-8mm}
\\
\includegraphics[scale=0.62]{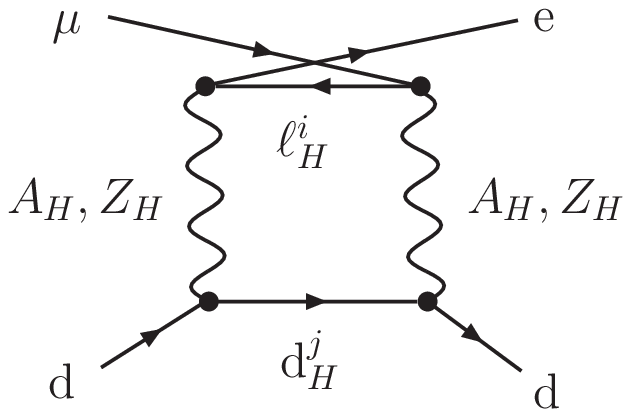} & \hspace{-8mm}
\includegraphics[scale=0.62]{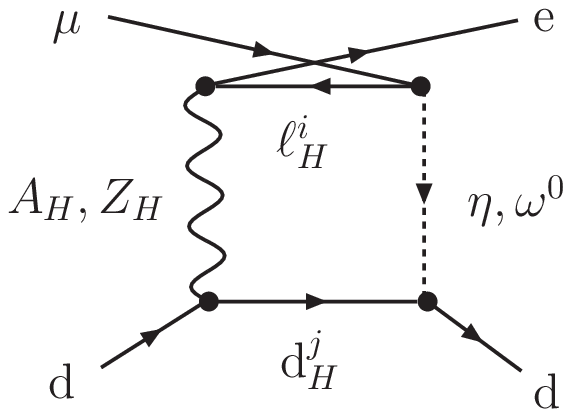} & \hspace{-8mm}
\includegraphics[scale=0.62]{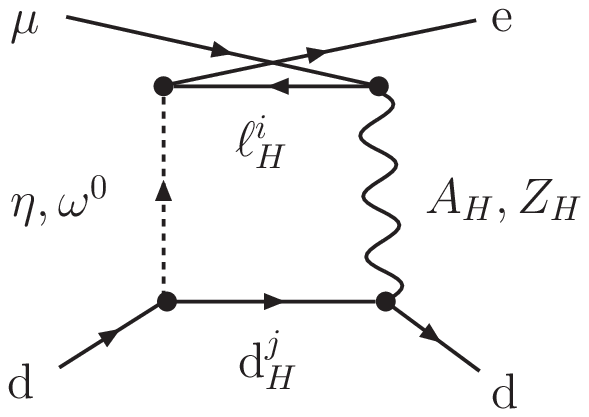} & \hspace{-8mm}
\includegraphics[scale=0.62]{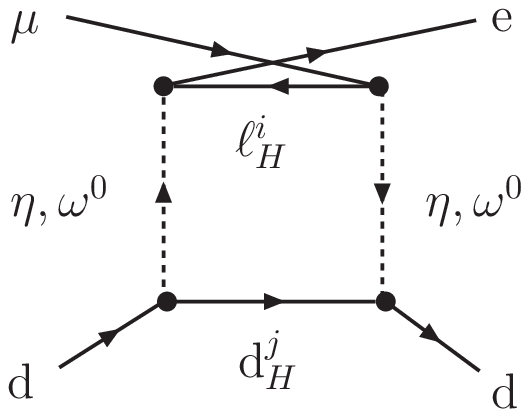}   \hspace{-8mm}
\\
\includegraphics[scale=0.62]{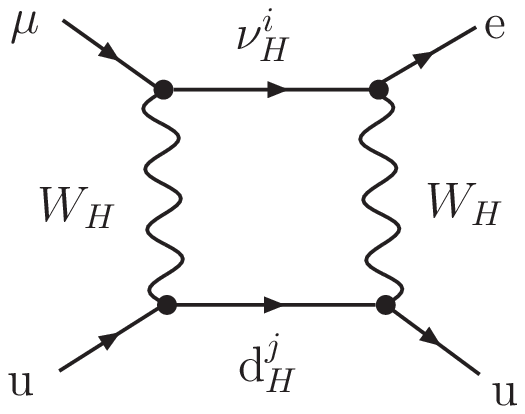}   & \hspace{-8mm}
\includegraphics[scale=0.62]{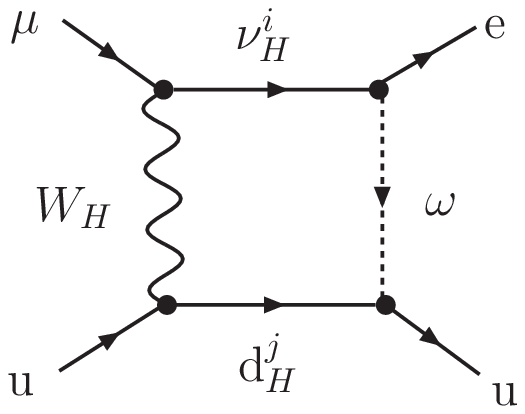}   & \hspace{-8mm}
\includegraphics[scale=0.62]{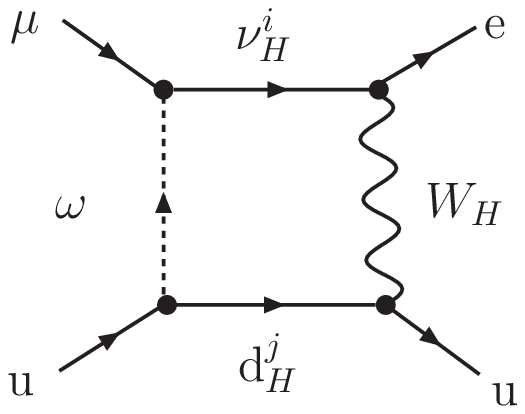}   & \hspace{-8mm}
\includegraphics[scale=0.62]{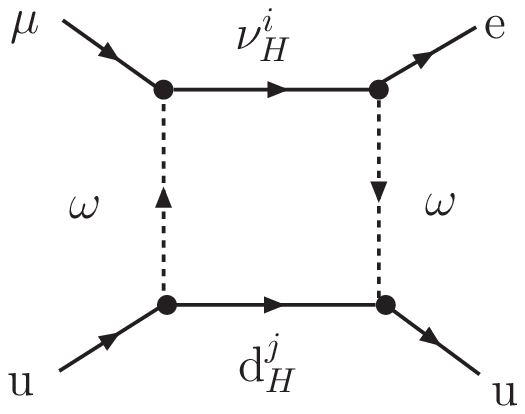}     \hspace{-8mm}
\\
\includegraphics[scale=0.62]{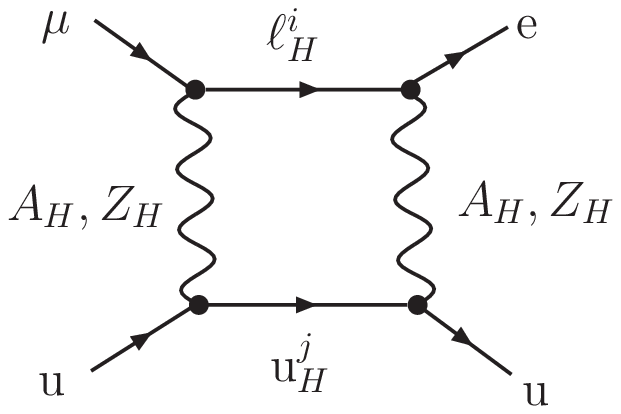} & \hspace{-8mm}
\includegraphics[scale=0.62]{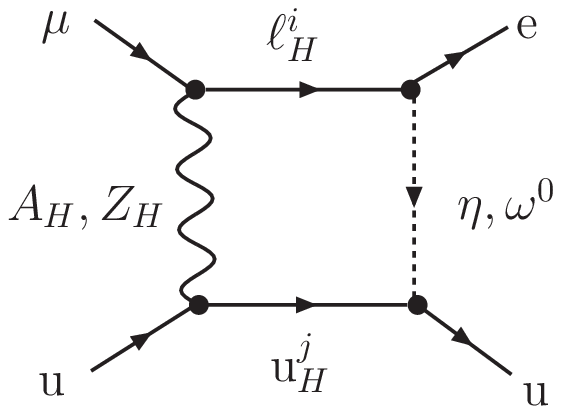} & \hspace{-8mm}
\includegraphics[scale=0.62]{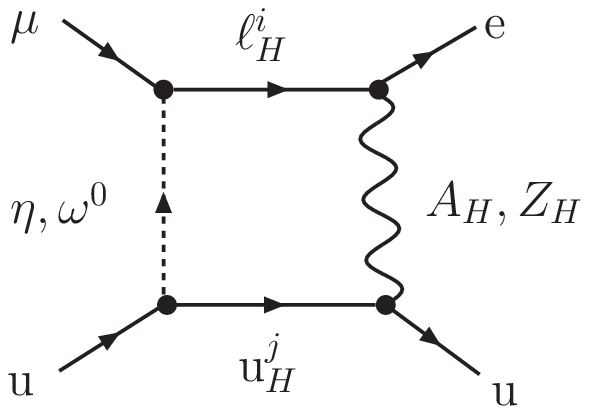} & \hspace{-8mm}
\includegraphics[scale=0.62]{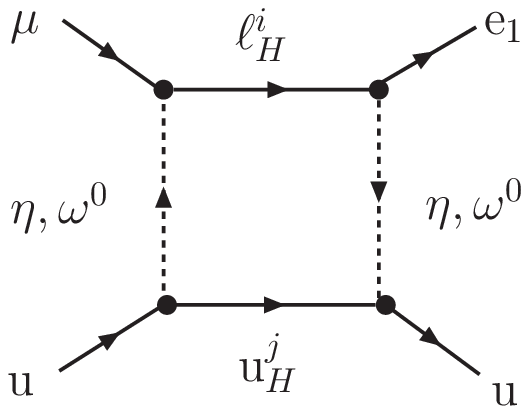}   \hspace{-8mm}
\\
\includegraphics[scale=0.62]{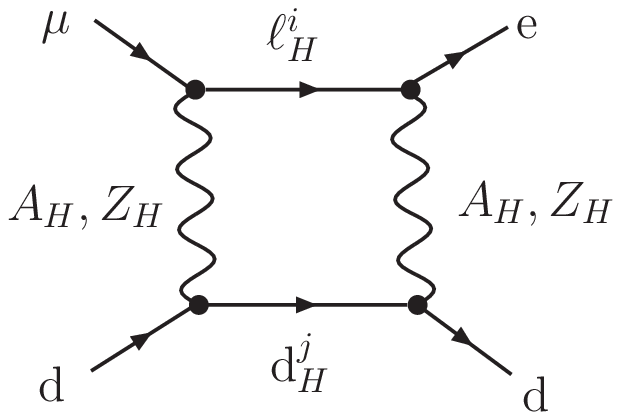} & \hspace{-8mm}
\includegraphics[scale=0.62]{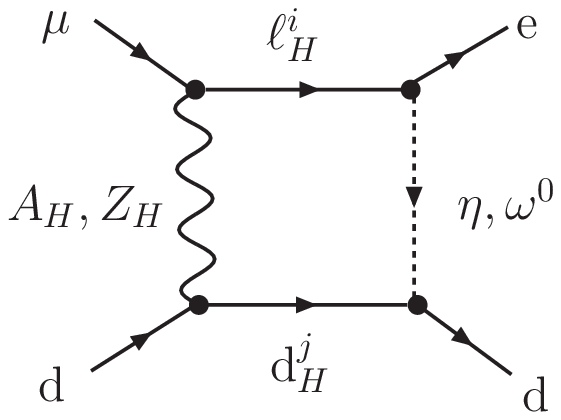} & \hspace{-8mm}
\includegraphics[scale=0.62]{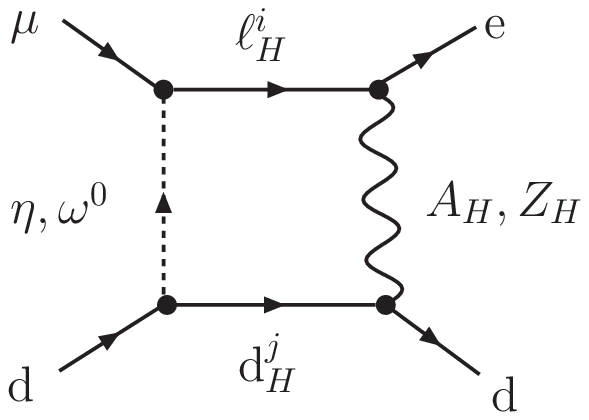} & \hspace{-8mm}
\includegraphics[scale=0.62]{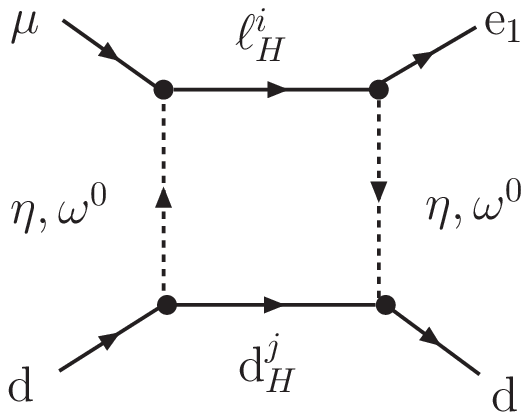}   \hspace{-8mm}
\end{tabular}
\caption{Box contributions to $\mueN$ in the LHT model.}
\label{boxes}
\end{figure}

The boxes do differ considerably and need to be recalculated. The list of diagrams is shown in figure~\ref{boxes}. The contributions are split into two form factors depending on the quark that enters on the lower legs:  $B_{1u}^L$ and $B_{1d}^L$. Notice that all diagrams with neutral gauge bosons include several contributions, i.e. diagrams with one or other type of boson and diagrams with mixed types.  The first diagram in the second row, for instance, includes four of them: $(A_H,A_H)$, $(Z_H,Z_H)$, $(A_H,Z_H)$, $(Z_H,A_H)$.

The resulting form factors are as follows:
\begin{equation}
\begin{split}
B_{1u}^L =\frac{\alpha_W}{16\pi}\frac{1}{s_W^2}\frac{1}{M_W^2}\frac{v^2}{4f^2}&\sum_{ij}\chi_{ij}^u
\Bigg[-\left(4+\frac{1}{4}y_iy_j^d\right)\widetilde d_0(y_i,y_j^d)
+2y_iy_j^d d_0(y_i,y_j^d)\\
&-\frac{3}{4}\widetilde{d}_0(y_i,y_j^u)-\frac{3}{100a}\widetilde{d}_0(ay_i,ay_j^u)+\frac{3}{10}\widetilde{d}_0(a,ay_i,ay_j^u)\Bigg],
\end{split}
\end{equation}
\begin{equation}
\begin{split}
B_{1d}^L =\frac{\alpha_W}{16\pi}\frac{1}{s_W^2}\frac{1}{M_W^2}\frac{v^2}{4f^2}&\sum_{ij}\chi_{ij}^d
\Bigg[\left(1+\frac{1}{4}y_iy_j^u\right)\widetilde d_0(y_i,y_j^u)-2y_iy_j^u d_0(y_i,y_j^u)\\
&-\frac{3}{4}\widetilde{d}_0(y_i,y_j^d)-\frac{3}{100a}\widetilde{d}_0(ay_i,ay_j^d)-\frac{3}{10}\widetilde{d}_0(a,ay_i,ay_j^d)\Bigg].
\end{split}
\end{equation}
Here we have introduced
\begin{eqnarray}
\chi_{ij}^{u}=V_{H\ell}^{i\mu}V_{H\ell}^{ie*}|V_{Hu}^{ju}|^2, \qquad
\chi_{ij}^{d}=V_{H\ell}^{i\mu}V_{H\ell}^{ie*}|V_{Hd}^{jd}|^2,
\end{eqnarray}
and
\begin{eqnarray}
y_i^u  = \frac{m_{u_H^i}^2}{M_{W_H}^2}, \quad y_i^d  = \frac{m_{d_H^i}^2}{M_{W_H}^2}.
\end{eqnarray}
The box functions $d_0(x,y,z)$, $\widetilde{d}_0(x,y,z)$, $d_0(x,y)$ and $\widetilde{d}_0(x,y)$ are given in equations (C.26) to (C-29) of \cite{delAguila:2008zu}.

\section{Numerical Results}

Next we can proceed to compute the conversion rates for different values of the LHT parameters.  As in our previous work \cite{delAguila:2008zu}, we restrict ourselves to the case of two lepton generations.  Then, the lepton sector can be fully determined by four parameters:  The LH breaking scale $f$, the masses of the heavy leptons\footnote{
We neglect the difference between the masses of the heavy neutrino and the charged lepton of the same generation, that is of order of $(v/f)^2$, as for quarks in (\ref{mH}).}
 $m_{Hi}$ $(i=1,2)$, and the angle $\theta$ that defines the now $2\times 2$ mixing matrix in the lepton sector $V_{H\ell}$:
\begin{equation}
 V_{H\ell} =  \left(\begin{array}{cc} V_{H\ell}^{1e} & V_{H\ell}^{1\mu} \\ V_{H\ell}^{2e} & V_{H\ell}^{2\mu} \end{array} \right) = 
\left( \begin{array}{cc} \cos \theta & \sin \theta \\ -\sin \theta & \cos \theta \end{array}\right)
\end{equation}
We shall again replace the masses $m_{Hi}$ by the parameters $\delta$ and $\tilde{y}$ defined as:
\begin{gather}
 \tilde{y}=\sqrt{y_1y_2}, \quad y_i = \frac{m_{Hi}^2}{M_{W_H}^2}, \quad i=1,2, \\
 \delta = \frac{m_{H2}^2-m_{H1}^2}{m_{H1}m_{H2}}.
\end{gather}
Both $m_{Hi}$ and $M_{W_H}$ are proportional to $f$ and, therefore, $\tilde{y}$ and $\delta$ are independent of this scale.

All form factors for this process then take the following general form:
\begin{equation}
 \sum_{i=1,2} V_{H\ell}^{ie*}V_{H\ell}^{i\mu} F(y_i) = \frac{\sin{2\theta}}{2} [F(y_1) - F(y_2)],
\end{equation}
with $F$ a generic function.  The dependence of the conversion rate for small $\delta$ can be approximated by:
\begin{equation}
 \mathcal{R} \propto \left|\frac{v^2}{f^2} \, \delta \sin{2\theta} \right|^2.
\end{equation}
Here the dependence on $f$ and $\theta$ is exact.  There is, of course, a dependence in $\tilde{y}$ but the behaviour with changes in this parameter cannot be expressed as simply. 

The quark sector requires another set of masses and mixings.  The two mixing matrices involved in the quark sector, $V_{Hu}$ and $V_{Hd}$, are related by $V_{Hd} = V_{Hu}V_{\rm CKM}$ so there is, in fact, only one matrix that requires fixing.  We will assign masses to the three heavy quarks ignoring the small mass difference (\ref{mH}) between up- and down-type quarks of the same generation ($y_i^u\approx y_i^d$).  The parameters for this sector are defined analogously to the three-family lepton sector in \cite{delAguila:2008zu}:
\begin{gather}
 \tilde{y}^u=\sqrt{y^u_1y^u_2}, \quad y^u_i = \frac{m_{u_H^i}^2}{M_{W_H}^2}, \quad i=1,2,3, \\
 \delta^u_{12} = \frac{m_{u_H^2}^2-m_{u_H^1}^2}{m_{u_H^1}m_{u_H^2}}, \quad \delta^u_{23} = \frac{m_{u_H^3}^2-m_{u_H^2}^2}{m_{u_H^2}m_{u_H^3}}. \label{quarkparameters}
\end{gather}

\subsection{Case without quark mixing}

\begin{figure}
\centering
 \begin{tabular}{cc}
  \includegraphics[height=.45\linewidth]{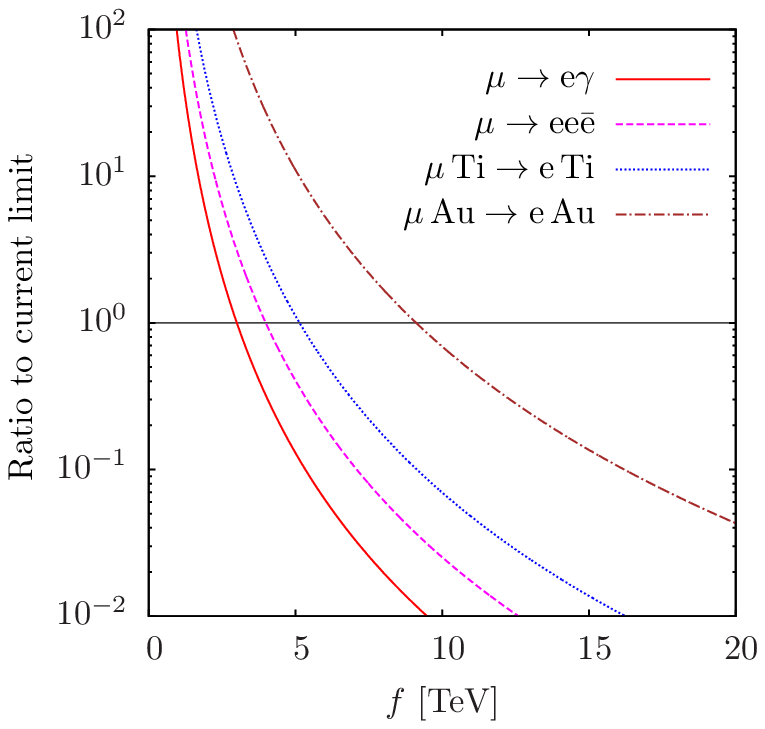} & \includegraphics[height=.45\linewidth]{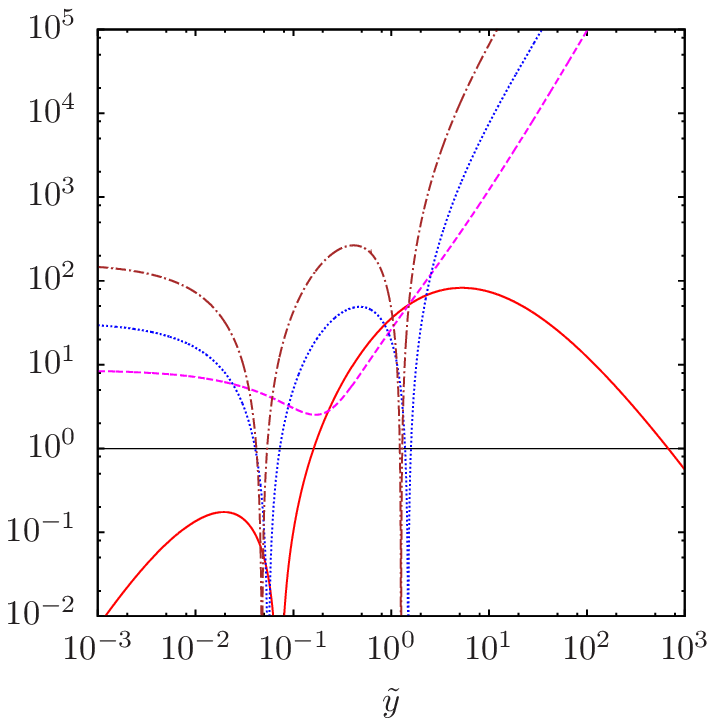} \\
  \includegraphics[height=.45\linewidth]{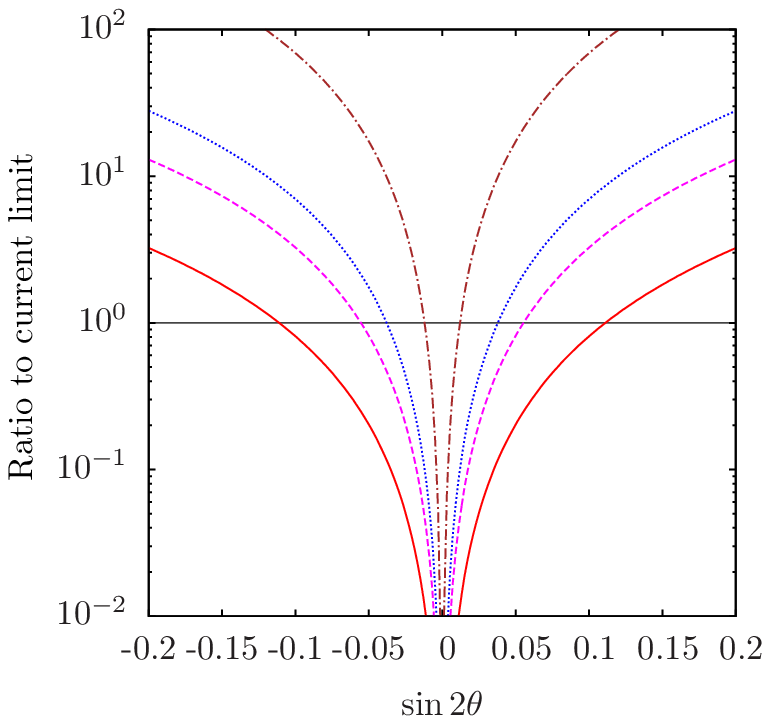} & \includegraphics[height=.45\linewidth]{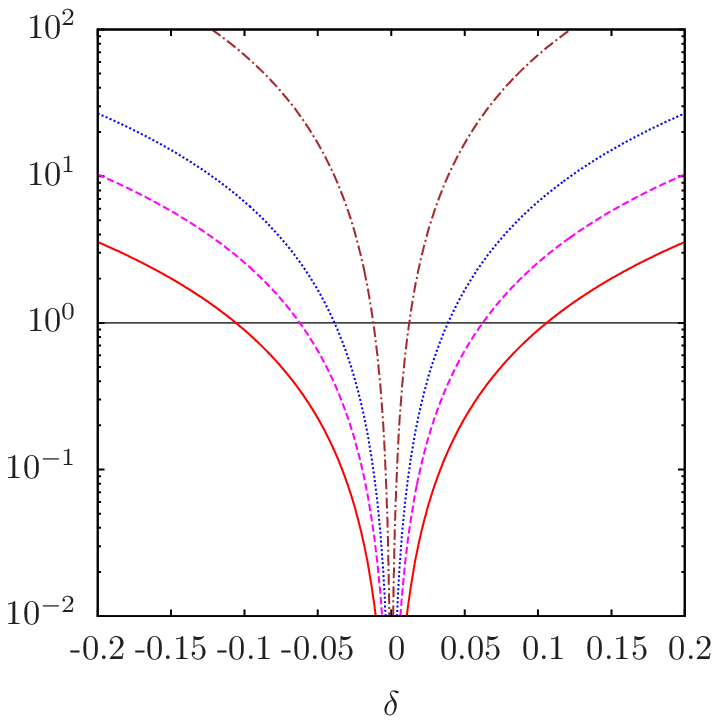}
 \end{tabular}
\caption{Ratios of LHT predictions to current limits, assuming $\tilde y^u=1$ and degenerate heavy quark masses, as functions of the different parameters keeping the others at their reference values.}
\label{plot1}
\end{figure}

We will firstly assume no flavor mixing in the quark sector, which can be achieved by assuming mass degeneracy of heavy quarks.  This is convenient to view the general behavior of the model clearly.  In figure~\ref{plot1} we show the conversion rates divided by the current experimental limits ($\mathcal{R}(\mu\, {\rm Au} \to {\rm e\, Au}) < 7 \times 10^{-13}$ \cite{Bertl:2006up} and $\mathcal{R}(\mu\, {\rm Ti} \to {\rm e\, Ti}) <4.3 \times 10^{-12}$ \cite{Dohmen:1993mp}).  Only values below unity are experimentally allowed.  The results of \cite{delAguila:2008zu} for $\muegamma$ ($\mathcal{B} < 1.2 \times 10^{-11}$ \cite{Brooks:1999pu}) and $\mueee$ ($\mathcal{B} < 10^{-12}$ \cite{Bellgardt:1987du}) are also plotted using the same inputs. The masses for the heavy quarks have been chosen to correspond to $\tilde{y}^u = 1$ and $\delta_{12}^u = \delta_{23}^u = 0$. {\em Natural} values for $f=1$~TeV, lepton mixing $\theta=\pi/4$, mass splitting $\delta=1$ and $\tilde y=4$ are taken as a reference, here and in the following unless stated otherwise.  
Notice that the reference value for $\tilde y$ has been shifted from the one used in \cite{delAguila:2008zu}. The reason for this can be seen in the top right hand graph of figure \ref{plot1} which shows the dependence on $\tilde{y}$.  The value $\tilde{y} = 1$ chosen in \cite{delAguila:2008zu} falls very close to a point where penguin and box diagrams cancel each other out.  This artificially reduces the value of the conversion rate to values below the values expected for $\mueee$ and $\muegamma$ even though in almost any other region the $\mueN$ process dominates.  These cancellations also occur elsewhere in the parameter space and in the other processes.  This only means that the process in question does not set a bound on the parameters in that area.  The key observation is that the rates for the three processes, $\muegamma$, $\mueee$ and $\mueN$, in general do not simultaneously vanish for any of the values of the parameters making it impossible for the model to fit into those areas. Scanning over parameter space has been discussed in some detail previously \cite{Blanke:2007db,Blanke:2009am}, finding large correlations for non-vanishing cross-sections between different processes. Here we prefer to look for the parameter regions where all processes get small, fulfilling all experimental constraints. This is so when the heavy Yukawa couplings effectively align with those of the charged leptons, i.e. small $\theta$ or $\delta$. Obviously, all new contributions also cancel for a large new physics scale, since the corresponding cross-sections scale like $f^{-4}$.

Barring the possibility of these cancellations, it is clear from figure~$\ref{plot1}$ that the most restrictive process is $\mu\, {\rm Au}\to \rm{e\, Au}$,  whose constraints are somewhat more demanding than those previously obtained in \cite{delAguila:2008zu} for $\muegamma$ and $\mueee$ and than the ones for $\mu\, {\rm Ti}\to \rm{e\, Ti}$.  However, this last process is expected to have the greatest improvements in future experiments \cite{Ritt:2006cg,Mori:2007zza,Kuno:2005mm}.  We compare the bounds on the parameters derived from $\muegamma$ and $\mueee$ with the new ones from $\mueN$ in table~\ref{compare} for current and future measurements.  The bounds on each of the parameters come from keeping all the others at the reference values and finding the region where the conversion rate (or branching ratio) is within the experimental limit.  Notice that these bounds depend strongly on the choice of input values so these bounds are not strict.

\begin{table}
\centering
\begin{tabular}{|r|r|r|r|r|r|r|r|}
\hline
& \multicolumn{2}{c|}{$\muegamma$} & 
  \multicolumn{2}{c|}{$\mueee$} &
   $\mu\, {\rm Au}\to\e\, {\rm Au}$ &
  \multicolumn{2}{c|}{$\mu\, {\rm Ti}\to\e\, {\rm Ti}$} \\
\hline
\multicolumn{1}{|c|}{Limit}
& $1.2\times10^{-11}$ & $10^{-13}$ & 
  \quad $10^{-12}$          &  $10^{-14}$ &
   $7\times10^{-13}$ &
  $4.3\times10^{-12}$   &  $10^{-18}$ 
\\
\hline
$f/\mbox{TeV}>$ & 3.00 & 9.93 & 3.98 & 12.6 & 9.11 & 5.13  & 234 \\
$\sin2\theta<$  & 0.111 & 0.010 & 0.055 & 0.006 & 0.012 & 0.038 & $<10^{-4}$ \\
$|\delta|<$     & 0.106 & 0.010 & 0.062 & 0.006 & 0.012 & 0.039 & $<10^{-4}$ \\
\hline
\end{tabular}
\caption{Constraints on LHT parameters from present and future experimental exclusion limits, assuming $\tilde y^u=1$ and degenerate heavy quark masses.}
\label{compare}
\end{table}

\begin{figure}
\centering
 \begin{tabular}{c}
  \includegraphics{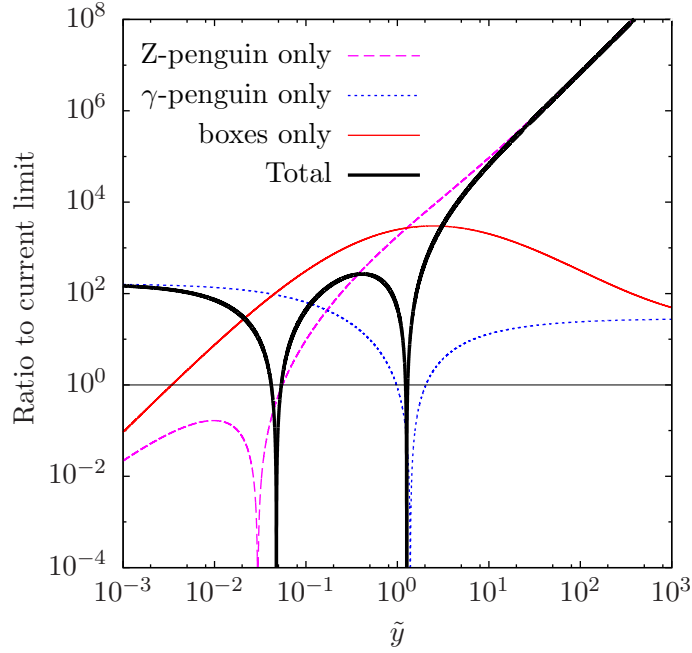}
 \end{tabular}
\caption{Contributions from photon penguins, Z penguins, boxes and their coherent sum to the total $\mu-\e$ conversion rate in Au, assuming $\tilde y^u=1$, degenerate heavy quark masses and reference values for the remaining parameters.}
\label{largeytilde}
\end{figure}

A comment about the large $\tilde y$ behavior is in order. Figure~\ref{largeytilde} shows the contributions of photon penguins, Z penguins, boxes and their coherent sum to the total $\mu-\e$ conversion rate in Au. Notice that only the $Z$ penguins are responsible for the non-decoupling effect, similar to that of the top quark in the SM \cite{Sirlin:1980nh,Hollik:1988ii,Bernabeu:1987me}. They saturate the $\mu-\e$ conversion rate for large $\tilde y$, analogously as for $\mueee$ in figure~\ref{plot1}. For such large values the main contributions results from (\ref{Fpenguin}), which grows like $\kappa_1\kappa_2$ for large heavy Yukawa coupling $\kappa_i$, $\tilde y=\sqrt{y_1y_2}\propto \kappa_1\kappa_2$.

\begin{figure}
 \centering
  \hspace{-0.5cm}\includegraphics[width=\linewidth]{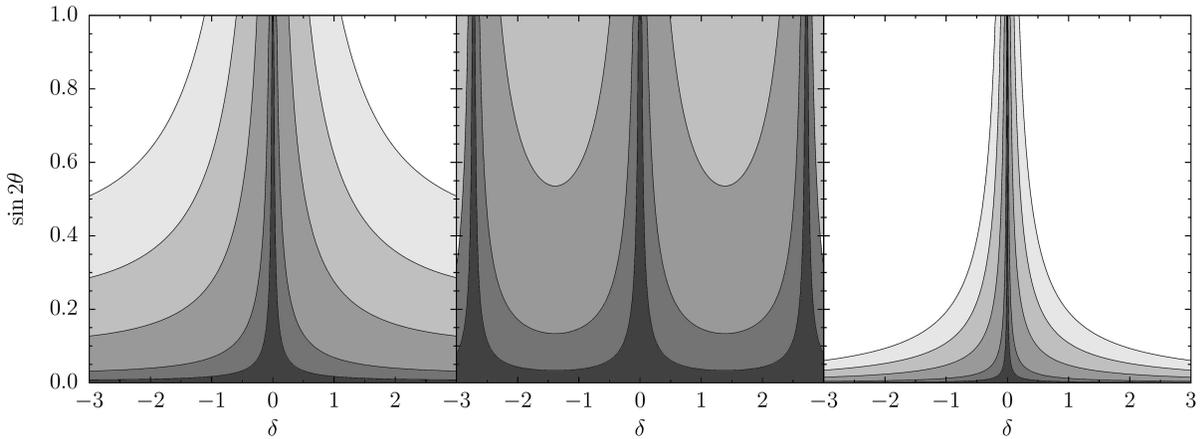}
\caption{Contours of $\mathcal{R}(\mu\, {\rm Au} \to \rm{e\, Au}) = 4.3\times 10^{-12}$ for values of $f = 0.5$, 1, 2, 3, 4 TeV (from bottom up) and $\tilde{y} = 0.25,$ 1, 4 (left to right), assuming $\tilde y^u=1$ and degenerate heavy quark masses.}
\label{contours}
\end{figure}

Figure~\ref{contours} shows exclusion contours in the $(\sin 2\theta,\delta)$ plane for the present experimental constraints for $\mu\, {\rm Au} \to e\, {\rm Au}$.  The case of $\tilde{y} = 1$ shows large allowed areas due to the fact that this is close to a cancellation point.  In the other cases, the mixing angle and the mass splitting are correlated if we are to remain within the experimental bound, as they were for $\muegamma$ and $\mueee$.  The bound is relaxed for higher values of the scale $f$.

\begin{figure}
 \centering
  \includegraphics[height=.45\linewidth]{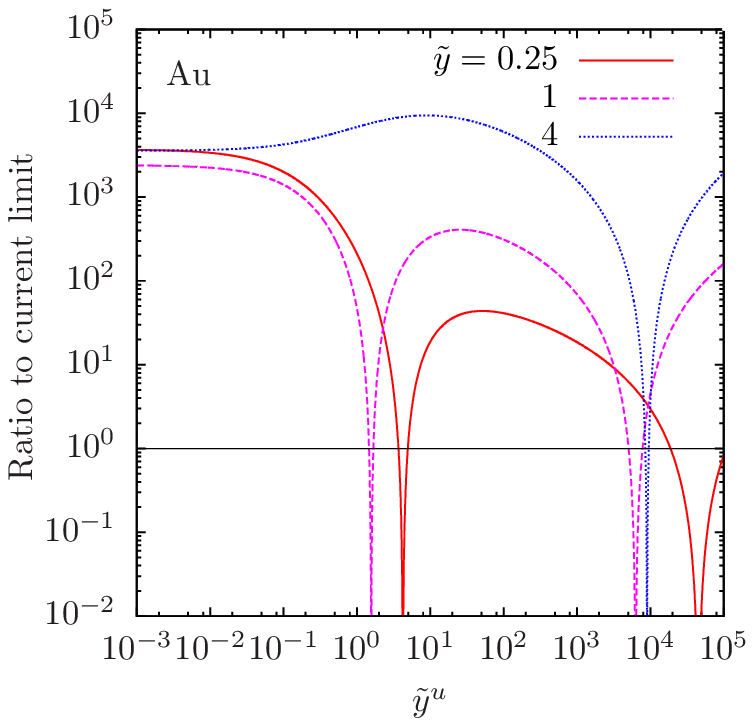} \includegraphics[height=.45\linewidth]{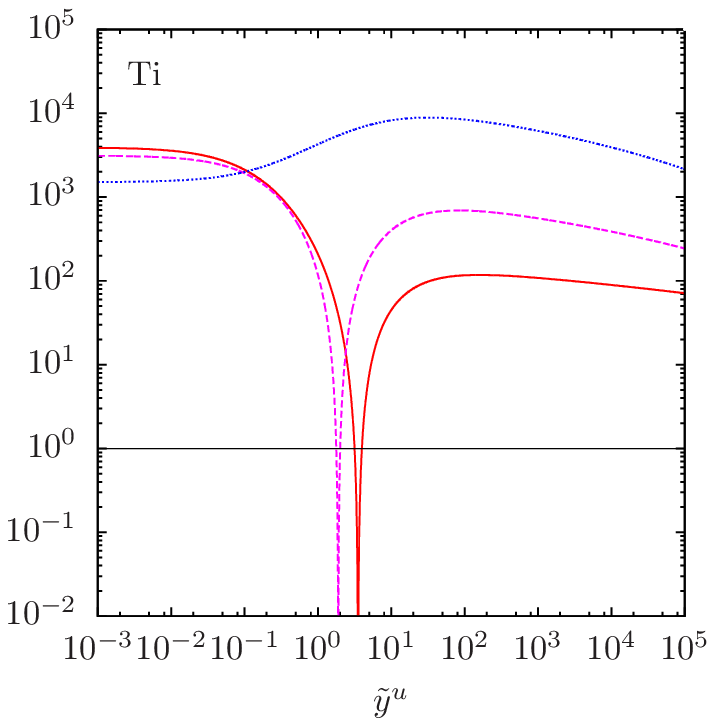}
\caption{Dependence of the $\mueN$ conversion rate on the heavy quark mass parameter $\tilde{y}^u$ for degenerate heavy quarks and several values of the lepton mass parameter $\tilde{y}$.}
\label{qtu}
\end{figure}

In figure~\ref{qtu} we show the dependence on the heavy quark mass parameter $\tilde{y}^u$ while keeping all quark masses degenerate.  We observe that the location of the cancellations depends strongly on the value of the lepton mass parameter $\tilde{y}$.  Because of these regions, even a degenerate heavy quark sector can be tuned to suppress the lepton flavor changing effect.  We therefore conclude that the values of the quark masses can have large effects on the conversion rate. The regions themselves, however, are relatively narrow concentrating on very definite $\tilde{y}^u$ values.

\subsection{Quark mixing effects}

\begin{figure}
\centering
 \begin{tabular}{cc}
  \includegraphics[height=.45\linewidth]{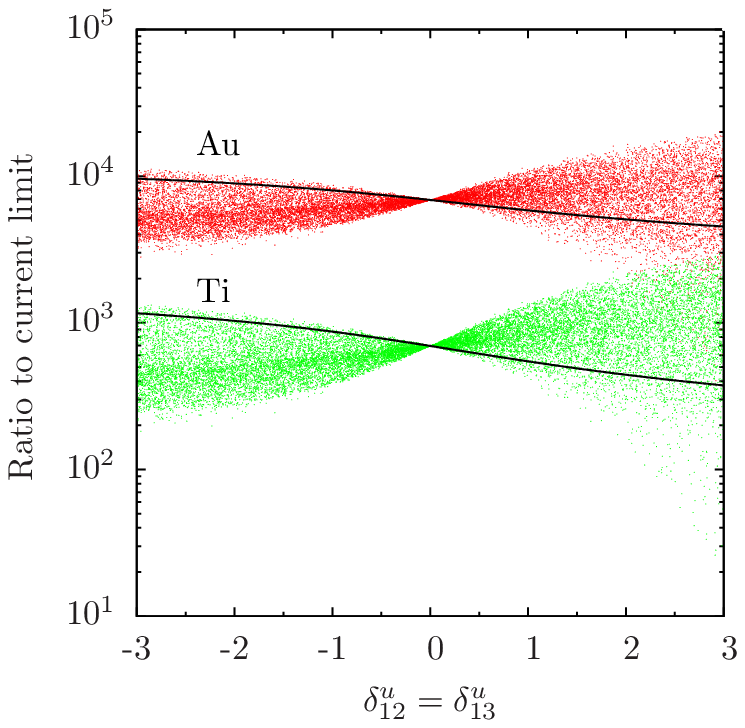} & \includegraphics[height=.45\linewidth]{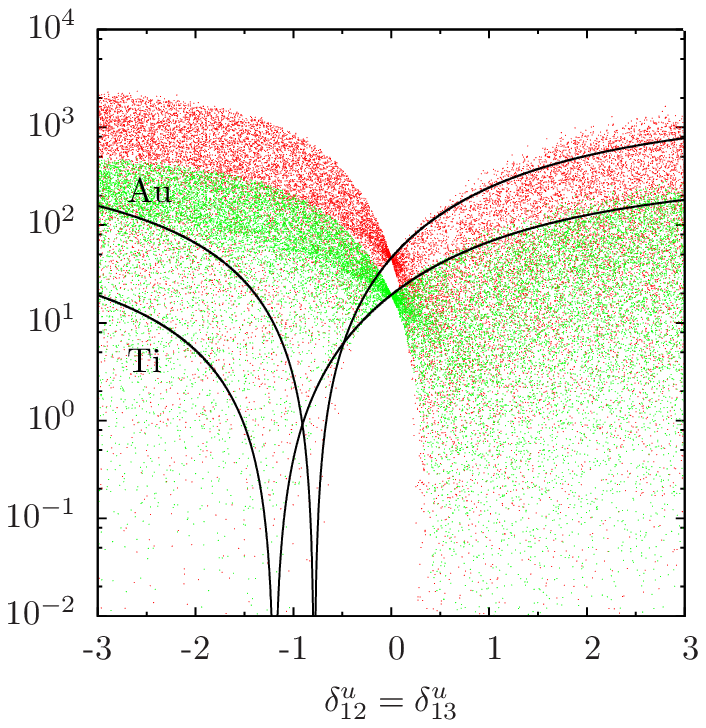}
 \end{tabular}
\caption{Ratio of conversion rate to current limit as a function of the heavy quark mass splittings for random values of mixing angles.  The solid lines correspond to $V_{Hd}=V_{Hu}V_{\rm CKM}=V_{\rm CKM}$. The average quark mass is fixed to $\tilde{y}^u = 1$ and the lepton mass parameter to $\tilde{y}=4$ (left) and $\tilde{y}=1$ (right).  Red (green) points are for Au (Ti).}
\label{scatterdelta}
\end{figure}

To study the effect on the conversion rate of the mixing matrix as it compares to the effect of the quark masses, we must choose the quark mass parameters in an area free of accidental suppressions.  One such choice is the quark mass parameter $\tilde{y}^u = 1$ and the lepton mass parameter $\tilde{y}=4$.  With these values, in figure~\ref{scatterdelta} we make a scatter plot showing the conversion rate as a function of the quark mass splittings.  The quark mixing angles and phases in $V_{Hu}$ ($V_{Hd}=V_{Hu}V_{\rm CKM}$) are uniformly distributed in their full range.  The solid lines correspond to minimal heavy quark mixing, that is, alignment of heavy and light quark flavors ($V_{Hu}$ = I and $V_{Hd}=V_{\rm CKM}$).

From the right panel of figure~\ref{scatterdelta} we can conclude that the mixing angles, although numerically important, do not alter the result as much as the mass parameters $\tilde{y}^u$, $\delta_{12}^u$ and $\delta_{23}^u$, as long as we are far from a cancellation area.  However, taking $\tilde{y} = 1$ the right panel shows that one can obtain virtually any value for the conversion rate by choosing the mixing angles appropriately.

\section{Conclusions}

As previous studies have demonstrated, LHT models are heavily constrained by flavor.  In this paper we complete our study presented in \cite{delAguila:2008zu} with the third basic LFV process, $\mueN$.  We have complemented other calculations \cite{Blanke:2007db,Choudhury:2006sq} of lepton flavor violating effects for this process and compared with predictions for $\muegamma$ and $\mueee$.

We have first briefly reviewed the quark sector Lagrangians for the LHT and find agreement with Feynman rules calculated in \cite{Blanke:2006eb}.  Generic limits on the LHT parameters can be found in table \ref{compare}.  In earlier works only conversion in Ti was considered but we find that, currently, conversion in Au gives the most stringent limits on the LHT parameters in general, producing normalized conversion rates (i.e. the conversion rate divided by the current experimental limit) of up to an order of magnitude larger than in the Ti case and requiring a Little Higgs breaking scale of almost 10 TeV or fine tuning at the percent level of the mixing angles and mass splittings in the lepton sector, which must be correlated in order to fulfill current experimental limits.  However, future limits on the Ti conversion rate will clearly be the most restrictive.

Nevertheless, certain values of the different parameters can suppress the conversion rate by causing accidental cancellations among the different contributions.  This means that, in some small regions, these processes do not actually provide any restrictions on the model, and only those derived from the other processes remain.  It is important to note, however, that the cancellation regions for the various processes do not in general overlap and therefore do not allow for the model to survive within them, since there is always at least one process that is above the experimental limits.  For $\mueN$, the origin of these cancellations is typically a sign difference between the penguin contributions and boxes, which allows them to cancel each other.

We have also considered the effect of a more general quark sector where we include non-degenerate quark masses and arbitrary mixings, whereas previous studies assumed a degenerate quark sector.  We find that even in the degenerate case, there is a sizable influence of the heavy quark mass parameter and, in some cases, it can completely cancel the conversion rate by pushing the prediction into one of the aforementioned suppression regions.  We also observe that the remaining quark sector parameters, namely mass splitting and mixing angles, can also shift the predictions somewhat although, as may be expected, the effect is smaller (less than a factor 10 in the normalized conversion rate) as long as we are far from a suppression region.  Otherwise, moving the angles and splittings can again accidentally reduce the conversion rate and change the value considerably or even cancel it completely.

\subsection*{Acknowledgments}
Work supported by the Spanish MICINN (FPA2006-05294), Junta de Andaluc{\'\i}a (FQM 101, FQM 03048) and European Community's Marie-Curie Research Training Network under contract MRTN-CT-2006-035505 ``Tools and Precision Calculations for Physics Discoveries at Colliders''.

\appendix

\section{Comparison with previous calculations}

We present below the relation of the form factors for $\muegamma$, $\mueee$ and $\mueN$ used in this and our previous work \cite{delAguila:2008zu}, which follow from \cite{Hisano:1995cp,Arganda:2005ji}, with those introduced by Buras et al. \cite{Blanke:2006eb,Blanke:2007db} corrected in \cite{Blanke:2009am}:
\bea
\frac{G_F}{2\sqrt{2}\pi^2}\bar D^{\prime\mu\e}_{\rm e,odd}&=&2A_2^R,
\label{a1}\\
\frac{G_F}{2\sqrt{2}\pi^2}\bar Z^{\mu\e}_{\rm odd}&=&-\frac{1}{2}(A_1^L+F_{LR}),
\label{a2}\\
\frac{G_F}{2\sqrt{2}\pi^2}\bar Y^{\mu\e}_{e,{\rm odd}}&=&\frac{s^2_W}{2}(B_1^L+2F_{LL}-2F_{LR}),
\label{a3}\\
\frac{G_F}{2\sqrt{2}\pi^2}\bar X^{\mu\e}_{\rm odd}&=&-s_W^2(B_{1u}^L+F_{LL}^u-F_{LR}^u),
\label{a4}\\
\frac{G_F}{2\sqrt{2}\pi^2}\bar Y^{\mu\e}_{\rm odd}&=&s_W^2(B_{1d}^L+F_{LL}^d-F_{LR}^d).
\label{a5}\eea

In the comparison with their results for the final expressions we have found agreement. Nevertheless, left and right hand sides of (\ref{a3}--\ref{a5}) differ by a common function of the heavy fermion mass involved in the line with no flavor change between external legs,
\bea
\frac{G_F}{2\sqrt{2}\pi^2}\sum_{ij}\chi_{ij} \left[\frac{7}{1-y_j}+\frac{y_j^2(8-y_j)\ln y_j}{(1-y_j)^2}\right],
\eea
that vanishes due to the unitarity of the mixing matrices,
\bea
\sum_{i}\chi_{ij}=0.
\eea 



\begin{thebibliography}{99}

{\small 

\bibitem{Mohapatra:1998rq}
  For a pedagogical introduction, see: R.~N.~Mohapatra and P.~B.~Pal,
  {\em Massive neutrinos in physics and astrophysics. Second edition},
  World Sci.\ Lect.\ Notes Phys.\  {\bf 60} (1998) 1
  [World Sci.\ Lect.\ Notes Phys.\  {\bf 72} (2004) 1].

\bibitem{Buras:2009if}
  A.~J.~Buras,
  {\em flavor Theory: 2009},
  arXiv:0910.1032 [hep-ph].

\bibitem{PDG}
  C.~Amsler {\it et al.}  [Particle Data Group],
  {\em Review of particle physics},
  Phys.\ Lett.\  B {\bf 667} (2008) 1 and
 2009 partial update for the 2010 edition.

\bibitem{Ritt:2006cg}
  S.~Ritt  [MEG Collaboration],
  {\em Status of the MEG experiment $\mu \to e \gamma$},
  Nucl.\ Phys.\ Proc.\ Suppl.\  {\bf 162} (2006) 279.

\bibitem{Mori:2007zza}
  T.~Mori,
  {\em MEG: The experiment to search for $\mu \to e \gamma$},
  Nucl.\ Phys.\ Proc.\ Suppl.\  {\bf 169} (2007) 166.

\bibitem{Kuno:2005mm}
  Y.~Kuno,
  {\em PRISM/PRIME},
  Nucl.\ Phys.\ Proc.\ Suppl.\  {\bf 149} (2005) 376.

\bibitem{Bona:2007qt}
  M.~Bona {\it et al.},
  {\em SuperB: A High-Luminosity Asymmetric $e^{+} e^{-}$ Super Flavor Factory.
  Conceptual Design Report},
  arXiv:0709.0451 [hep-ex].

\bibitem{ArkaniHamed:2001ca}
  N.~Arkani-Hamed, A.~G.~Cohen and H.~Georgi,
  {\em (De)constructing dimensions},
  Phys.\ Rev.\ Lett.\  {\bf 86} (2001) 4757
  [arXiv:hep-th/0104005].

\bibitem{ArkaniHamed:2001nc}
  N.~Arkani-Hamed, A.~G.~Cohen and H.~Georgi,
  {\em Electroweak symmetry breaking from dimensional deconstruction},
  Phys.\ Lett.\  B {\bf 513} (2001) 232
  [arXiv:hep-ph/0105239].

\bibitem{Schmaltz:2005ky}
  M.~Schmaltz and D.~Tucker-Smith,
  {\em Little Higgs review},
  Ann.\ Rev.\ Nucl.\ Part.\ Sci.\  {\bf 55} (2005) 229
  [arXiv:hep-ph/0502182].

\bibitem{Han:2005ru}
  T.~Han, H.~E.~Logan and L.~T.~Wang,
  {\em Smoking-gun signatures of little Higgs models},
  JHEP {\bf 0601} (2006) 099
  [arXiv:hep-ph/0506313].

\bibitem{Perelstein:2005ka}
  M.~Perelstein,
  {\em Little Higgs models and their phenomenology},
  Prog.\ Part.\ Nucl.\ Phys.\  {\bf 58} (2007) 247
  [arXiv:hep-ph/0512128].

\bibitem{ArkaniHamed:2002qy}
  N.~Arkani-Hamed, A.~G.~Cohen, E.~Katz and A.~E.~Nelson,
  {\em The littlest Higgs},
  JHEP {\bf 0207} (2002) 034
  [arXiv:hep-ph/0206021].

\bibitem{Csaki:2002qg}
  C.~Cs\'aki, J.~Hubisz, G.~D.~Kribs, P.~Meade and J.~Terning,
  {\em Big corrections from a little Higgs},
  Phys.\ Rev.\  D {\bf 67} (2003) 115002
  [arXiv:hep-ph/0211124].

\bibitem{Csaki:2003si}
  C.~Csaki, J.~Hubisz, G.~D.~Kribs, P.~Meade and J.~Terning,
  {\em Variations of little Higgs models and their electroweak constraints},
  Phys.\ Rev.\  D {\bf 68} (2003) 035009
  [arXiv:hep-ph/0303236].

\bibitem{Han:2003wu}
  T.~Han, H.~E.~Logan, B.~McElrath and L.~T.~Wang,
  {\em Phenomenology of the little Higgs model},
  Phys.\ Rev.\  D {\bf 67} (2003) 095004
  [arXiv:hep-ph/0301040].

\bibitem{Kilian:2003xt}
  W.~Kilian and J.~Reuter,
  {\em The low-energy structure of little Higgs models},
  Phys.\ Rev.\  D {\bf 70} (2004) 015004
  [arXiv:hep-ph/0311095].

\bibitem{Hubisz:2005tx}
  J.~Hubisz, P.~Meade, A.~Noble and M.~Perelstein,
  {\em Electroweak precision constraints on the littlest Higgs model with T
  parity},
  JHEP {\bf 0601} (2006) 135
  [arXiv:hep-ph/0506042].

\bibitem{Chen:2006dy}
  M.~C.~Chen,
  {\em Models of little Higgs and electroweak precision tests},
  Mod.\ Phys.\ Lett.\  A {\bf 21} (2006) 621
  [arXiv:hep-ph/0601126].

\bibitem{Cheng:2003ju}
  H.~C.~Cheng and I.~Low,
  {\em TeV symmetry and the little hierarchy problem},
  JHEP {\bf 0309} (2003) 051
  [arXiv:hep-ph/0308199].

\bibitem{Cheng:2004yc}
  H.~C.~Cheng and I.~Low,
  {\em Little hierarchy, little Higgses, and a little symmetry},
  JHEP {\bf 0408} (2004) 061
  [arXiv:hep-ph/0405243].

\bibitem{Low:2004xc}
  I.~Low,
  {\em T parity and the littlest Higgs},
  JHEP {\bf 0410} (2004) 067
  [arXiv:hep-ph/0409025].

\bibitem{Blanke:2007db}
  M.~Blanke, A.~J.~Buras, B.~Duling, A.~Poschenrieder and C.~Tarantino,
  {\em Charged Lepton flavor Violation and $(g-2)_\mu$ in the Littlest Higgs Model with T-Parity: a clear Distinction from Supersymmetry},
  JHEP {\bf 0705} (2007) 013
  [arXiv:hep-ph/0702136].

\bibitem{delAguila:2008zu}
  F.~del \'Aguila, J.~I.~Illana and M.~D.~Jenkins,
  {\em Precise limits from lepton flavor violating processes on the Littlest
  Higgs model with T-parity},
  JHEP {\bf 0901}, 080 (2009)
  [arXiv:0811.2891 [hep-ph]].

\bibitem{Choudhury:2006sq}
  S.~R.~Choudhury, A.~S.~Cornell, A.~Deandrea, N.~Gaur and A.~Goyal,
  Phys.\ Rev.\  D {\bf 75} (2007) 055011
  [arXiv:hep-ph/0612327].

\bibitem{Blanke:2009am}
  M.~Blanke, A.~J.~Buras, B.~Duling, S.~Recksiegel and C.~Tarantino,
  {\em FCNC Processes in the Littlest Higgs Model with T-Parity: a 2009 Look},
  arXiv:0906.5454 [hep-ph].

\bibitem{Goto:2008fj}
  T.~Goto, Y.~Okada and Y.~Yamamoto,
  {\em Ultraviolet divergences of flavor changing amplitudes in the littlest Higgs model with T-parity},
  Phys.\ Lett.\  B {\bf 670} (2009) 378
  [arXiv:0809.4753 [hep-ph]].

\bibitem{Sirlin:1980nh}
  A.~Sirlin,
  {\em Radiative Corrections In The $SU(2)_L\times U(1)$ Theory: A Simple
  Renormalization Framework},
  Phys.\ Rev.\  D {\bf 22} (1980) 971. 

\bibitem{Hollik:1988ii}
  W.~F.~L.~Hollik,
  {\em Radiative Corrections in the Standard Model and their Role for Precision
  Tests of the Electroweak Theory},
  Fortsch.\ Phys.\  {\bf 38} (1990) 165.

\bibitem{Bernabeu:1987me}
  J.~Bernab\'eu, A.~Pich and A.~Santamar{\'\i}a,
  {\em $\Gamma(Z \to b\bar b)$: A Signature Of Hard Mass Terms For A Heavy Top},
  Phys.\ Lett.\  B {\bf 200} (1988) 569.

\bibitem{paper3}
  F.~del \'Aguila, J.~I.~Illana and M.~D.~Jenkins,	 
  work in preparation.

\bibitem{Illana:2009yp}
  J.~I.~Illana and M.~D.~Jenkins,
  {\em Lepton Flavor Violation in Little Higgs Models},
  Acta Phys.\ Polon.\  B {\bf 40} (2009) 3143
  [arXiv:0911.2173 [hep-ph]].

\bibitem{Agashe:2006iy}
  K.~Agashe, A.~E.~Blechman and F.~Petriello,
  {\em Probing the Randall-Sundrum geometric origin of flavor with lepton
   flavor violation},
  Phys.\ Rev.\  D {\bf 74} (2006) 053011
  [arXiv:hep-ph/0606021].

\bibitem{delAguila:2010vg}
  F.~del Aguila, A.~Carmona and J.~Santiago,
  {\em Neutrino Masses from an A4 Symmetry in Holographic Composite Higgs
  Models},
  arXiv:1001.5151 [hep-ph].

\bibitem{Csaki:2010aj}
  C.~Csaki, Y.~Grossman, P.~Tanedo and Y.~Tsai,
  {\em Warped Penguins},
  arXiv:1004.2037 [hep-ph].

\bibitem{Blanke:2006eb}
  M.~Blanke, A.~J.~Buras, A.~Poschenrieder, S.~Recksiegel, C.~Tarantino, S.~Uhlig and A.~Weiler,
  {\em Rare and CP-Violating $K$ and $B$ Decays in the Littlest Higgs Model with T Parity},
  JHEP {\bf 0701}, 066 (2007)
  [arXiv:hep-ph/0610298].

\bibitem{Hubisz:2004ft}
  J.~Hubisz and P.~Meade,
  {\em Phenomenology of the littlest Higgs with T-parity},
  Phys.\ Rev.\  D {\bf 71} (2005) 035016
  [arXiv:hep-ph/0411264].

\bibitem{Blanke:2006xr}
  M.~Blanke, A.~J.~Buras, A.~Poschenrieder, S.~Recksiegel, C.~Tarantino, S.~Uhlig and A.~Weiler,
  Phys.\ Lett.\  B {\bf 646} (2007) 253
  [arXiv:hep-ph/0609284].

\bibitem{Hubisz:2005bd}
  J.~Hubisz, S.~J.~Lee and G.~Paz,
  {\em The flavor of a little Higgs with T-parity},
  JHEP {\bf 0606} (2006) 041
  [arXiv:hep-ph/0512169].

\bibitem{Hisano:1995cp}
  J.~Hisano, T.~Moroi, K.~Tobe and M.~Yamaguchi,
  {\em Lepton-Flavor Violation via right-handed neutrino Yukawa couplings in
  the Supersymmetric Standard Model},
  Phys.\ Rev.\  D {\bf 53} (1996) 2442
  [arXiv:hep-ph/9510309].

\bibitem{Illana:2000ic}
  J.~I.~Illana and T.~Riemann,
  {\em Charged lepton flavor violation from massive neutrinos in Z decays},
  Phys.\ Rev.\  D {\bf 63} (2001) 053004
  [arXiv:hep-ph/0010193].

\bibitem{Illana:2002tg}
  J.~I.~Illana and M.~Masip,
  {\em Lepton flavor violation in Z and lepton decays in supersymmetric  models},
  Phys.\ Rev.\  D {\bf 67} (2003) 035004
  [arXiv:hep-ph/0207328].

\bibitem{Kuno:1999jp}
  Y.~Kuno and Y.~Okada,
  {\em Muon decay and physics beyond the standard model},
  Rev.\ Mod.\ Phys.\  {\bf 73}, 151 (2001)
  [arXiv:hep-ph/9909265].

\bibitem{Kitano:2002mt}
  R.~Kitano, M.~Koike and Y.~Okada,
  {\em Detailed calculation of lepton flavor violating muon electron  conversion rate for various nuclei},
  Phys.\ Rev.\  D {\bf 66} (2002) 096002
  [Erratum-ibid.\  D {\bf 76} (2007) 059902]
  [arXiv:hep-ph/0203110].

\bibitem{Bertl:2006up}
  W.~H.~Bertl {\it et al.}  [SINDRUM II Collaboration],
  {\em A Search for $\mu - e$ conversion in muonic gold},
  Eur.\ Phys.\ J.\  C {\bf 47} (2006) 337.

\bibitem{Dohmen:1993mp}
  C.~Dohmen {\it et al.}  [SINDRUM II Collaboration.],
  {\em Test of lepton flavor conservation in $\mu \to e$ conversion on Titanium},
  Phys.\ Lett.\  B {\bf 317} (1993) 631.

\bibitem{Brooks:1999pu}
  M.~L.~Brooks {\it et al.}  [MEGA Collaboration],
  {\em New limit for the family-number non-conserving decay $\mu^+ \to e^+ \gamma$},
  Phys.\ Rev.\ Lett.\  {\bf 83}, 1521 (1999)
  [arXiv:hep-ex/9905013].

\bibitem{Bellgardt:1987du}
  U.~Bellgardt {\it et al.}  [SINDRUM Collaboration],
  {\em Search for the decay $\mu^+ \to e^+ e^+ e^-$},
  Nucl.\ Phys.\  B {\bf 299}, 1 (1988).

\bibitem{Arganda:2005ji}
  E.~Arganda and M.~J.~Herrero,
  {\em Testing supersymmetry with lepton flavor violating tau and mu decays},
  Phys.\ Rev.\  D {\bf 73} (2006) 055003
  [arXiv:hep-ph/0510405].

}

\end{thebibliography}
\end{document}